\documentclass[12pt]{article}
\usepackage{amssymb,latexsym}
\usepackage[cmex10]{amsmath}
\usepackage[all]{xy}
\input{xy}
\xyoption{all}
\title{On Sequences, Rational Functions and Decomposition}
\author{Graham H. Norton\footnote{
School of Mathematics and Physics, University of Queensland, Brisbane, Queensland 4072, Australia (email: ghn@maths.uq.edu.au).}}
\newtheorem{theorem}{\sc Theorem}[section] 
\newtheorem{corollary}[theorem]{\sc Corollary} 
 \newtheorem{proposition}[theorem]{\sc Proposition} \newtheorem{definition}[theorem]{\sc Definition} 
\newtheorem{lemma}[theorem]{\sc Lemma} 
\newtheorem{example}[theorem]{\sc Example}  
\newtheorem{examples}[theorem]{\sc Examples}  
\newtheorem{algorithm}[theorem]{\sc Algorithm} 
\newtheorem{problem}[theorem]{\sc  Problem}
\newtheorem{remark}[theorem]{\sc  Remark}
\newtheorem{remarks}[theorem]{\sc  Remarks} 
\newtheorem{propdefn}[theorem]{\sc Proposition-Definition} 
\newenvironment{proof}{{\sc Proof.}}{\hspace*{\fill}$\square$\par\vspace{4mm}} 

\def \bt{ \begin{theorem} }
\def \et{ \end  {theorem} }
\def \bl{ \begin{lemma} }
\def \el{ \end  {lemma} }
\def \bp{ \begin{proposition} }
\def \ep{ \end  {proposition} }
\def \bc{ \begin{corollary} }
\def \ec{ \end  {corollary} }
\def \bd{ \begin{definition} }
\def \ed{ \end  {definition} }
\def \bpd{ \begin{propdefn} }
\def \epd{ \end  {propdefn} }
\def \bdt{ \begin{definitiontheorem} }
\def \edt{ \end  {definitiontheorem} }
\def \bpr{ \begin{proof} }
\def \epr{ \end  {proof} }
\def \ba{ \begin{algorithm} }
\def \ea{ \end{algorithm} }
\def \be{ \begin{example} }
\def \eex{ \end{example} }
\def \bes{ \begin{examples} }
\def \eexs{ \end{examples} }
\def \br{ \begin{remark} }
\def \er{ \end{remark} }
\def \brs{ \begin{remarks} }
\def \ers{ \end{remarks} }
\def \bpb{ \begin{problem} }
\def \epb{ \end{problem} }

\newcommand{\Ann} {\mathrm{Ann}}

\newcommand{\Min} {\mathrm{MP}}

\newcommand{\Pol} {\mathrm{Pol}}

\newcommand{\la} {\leftarrow}

\newcommand{\ul} {\underline}

\newcommand{\Id} {\mathrm{Id}}

\newcommand{\ra} {\rightarrow}

\newcommand{\vv} {\mathrm{v}}
\newcommand{\D} {\mathrm{D}}

\newcommand{\lc} {\mathrm{lc}}
\newcommand{\LC} {\mathrm{L}}

\newcommand{\N} {\mathbb{N}}

\newcommand{\R} {\mathrm{R}}

\newcommand{\F}{\mathbb{F}}

\newcommand{\Z}{\mathbb{Z}}

\begin{document}
\maketitle
\begin{abstract} 
It is classical that well-known identities and properties of  partial quotients furnish  rational approximation in $\F[[x^{-1}]]$. For a rational function, this is the extended Euclidean algorithm in $\F[x]$. Berlekamp's heuristic solution of the 'key equation' essentially approximates an element of $\F[x]$ with constant term 1 via a quotient of  reciprocals, and his solutions  satisfy a number of identities. In earlier papers we gave a solution of an analogous problem using $\D[x^{-1},x]$, $\D$ a commutative domain.  

The linear complexity (of a finite initial subsequence) of an infinite sequence over $\F$ has been related to the degrees of its partial quotients by Mills, Cheng, Niederreiter and others.
We use  first principles and induction to relate these linear complexities  to the degrees of its partial quotients.

 Berlekamp has also described the set of solutions of the key equation. 
We define a pairing of minimal solutions  and a 'minimal system' of a finite sequence over $\D$. Examples are classical approximation in $\F[[x^{-1}]]$ and approximation using $\D[x^{-1},x]$. We use minimal systems to generalise results of Massey and Niederreiter to arbitrary solutions, including numerators. This includes   explicit and unique decomposition of both parts of a solution into a sum of (polynomial) multiples of solutions with minimal degree denominators. The unique multipliers also satisfy degree constraints.  
 
We give several applications to gcd's of sequence polynomials and  relate  partial-quotient solutions to  solutions derived using $\F[x^{-1},x]$. We give a precise count of the number of solutions  when the field is finite. 
  Our final application concerns when the first component of a minimal solution vanishes at some scalar;  a simple modification of our approach gives a new solution, the first component of which does not vanish at the scalar and which has minimal degree. We also describe the corresponding    set of solutions.
  This simplifies and generalises work of Salagean.

We conclude that numerators (or second components) of solutions can play  a significant role in proofs of  properties of denominators (or first components) and that they enjoy similar properties.
\end{abstract}


\tableofcontents
\newpage
\section{Introduction}
\subsection{Background}
 Let $\F$ be a field. Approximating the generating function $S_1x^{-1}+S_2x^{-2}+\cdots\in x^{-1}\F[[x^{-1}]]$ by  rational functions  $q_2^{(i)}/q_1^{(i)}$ (where $i\geq 0$)  is well-known;  
 $q_1^{(i)}, q_2^{(i)}$ are known as its partial quotients or rational convergents. An important identity is 
\begin{eqnarray}\label{cfid}
q_2^{(i)}q_1^{(i-1)}- q_1^{(i)}q_2^{(i-1)}=(-1)^{i-1}.
\end{eqnarray}
Obtaining partial quotients uses division in the field of Laurent series in $x^{-1}$, written $\F((x^{-1}))$. 
When the above sum is a rational function, this is the extended Euclidean algorithm.  See also \cite{Mills} for connections with linear recurring sequences.

A second example  is Berlekamp's iterative solutions $\omega^{(i)}/\sigma^{(i)}$ of the 'key equation', where $0\leq i\leq n$, \cite[Section 7]{Be68}. (The integer $n$ is related to a decoding  problem.) It is essentially rational approximation of $1+s_1x+\cdots +s_nx^n\in\F[x]$ using reciprocals of polynomials. It uses 'auxiliary solutions' $\gamma^{(i)}/\tau^{(i)}$ which satisfy $\omega^{(i)}\tau^{(i)}-\sigma^{(i)}\gamma^{(i)}=x^i$,  \cite[Theorem 7.42]{Be68}. The set of  solutions was discussed in   \cite[Theorems 7.43, 7.44]{Be68}.

A simplification of Berlekamp's algorithm appeared in \cite[Algorithm 1]{Ma69}. This  interprets $\sigma^{(n)}$ as  a 'connection polynomial of a  minimal-length linear-feedback shift register (LFSR) which generates $s=s_1,\ldots,s_n$'. It is known as the Berlekamp-Massey algorithm. The minimal length is  called  the 'linear complexity' $\LC_n$ of  $s$.   The set of  connection polynomials for all LFSR's of length $\LC_n$ which generate $s$ was given in  \cite[Theorem 3]{Ma69}.

Connections between these two types of rational approximation e.g.  between the linear complexity of $S_1,\ldots,S_n$  and  the degrees of the denominators $q_1^{(i)}$ have been discussed in \cite{Cheng}, \cite{Mills} and \cite{WS},  which depend on \cite{Be68}. In \cite[Theorem 1]{Nied87} this was done  independently of \cite{Be68} and  \cite{Ma69}. 
  
 A third example appeared in \cite{N95b}. Our goal was a faithful  redevelopment and extension of \cite{Ma69}; we were unaware of \cite{Nied87} and Macaulay's inverse systems, see e.g. \cite{Northcott}. We discussed rational approximation of $s_0+\cdots+s_{1-n}x^{1-n}\in\D[x^{-1}]$  using  Laurent polynomials $\D[x^{-1},x]$, where $\D$ is a commutative domain. 
We write our  solution as  $\mu=(\mu_1,\mu_2)$ and call $\mu_1$ a 'minimal polynomial' of $s$. The linear complexity $\LC_n$ of $s$ is the degree of $\mu_1$ and the reciprocal of $\mu_1$ is a connection polynomial of an LFSR generating $s$, \cite{N10a}.  
When $\D$ is a field, our approach has applications to the above decoding problem  and to control theory,  see for instance \cite[Section 8]{N99b} and \cite[Example 4.9]{N95b}.

\subsection{Overview}
Our  overall goal is to unify and extend some results in the literature related to the rational  approximation of    generating functions of infinite and finite sequences. In our approach,  numerators play a significant role.

We  revisit  \cite[Theorem 1]{Nied87}, which has two parts. We give an inductive proof of the first part on  linear complexity and partial quotients. Our proof is  from first principles, using the basic definitions for finite sequences from \cite{N95b}. We also prove the converse.

We also derive an analogue of Identity (\ref{cfid}) for our minimal solutions, Proposition \ref{identity}. This enables us to 'decompose' solutions and determine the set of all solutions for a finite sequence over $\D$ whenever we have a 'minimal system' for $s$. Partial quotients (with $\D=\F$) also provide a minimal system. In this way we generalise the second part of \cite[Theorem 1]{Nied87} to all  solutions. 
We conclude with some applications of decomposition. 

{\em Note to the reader:}  we consider the partial quotients for an infinite sequence over  a field only; we have not extended \cite{Cheng} and \cite{Nied87} to commutative domains. 
In some situations, we apply our results to finite sequences over a field e.g.  Proposition \ref{minfield}, Proposition \ref{monic}  on monic minimal polynomials and Corollaries   \ref{minfieldconverse} -  \ref{Anncount}. 
We have included a number of examples; some  reappear intentionally in different guises as an expository aid and others  are inductive bases for later theorems.  
 \subsection{In More Detail}
We begin with basic concepts  for  infinite sequences over $\F$, denoted $S_0,S_{-1},\ldots$ and finite sequences over $\D$, denoted $s_0,\ldots,s_{1-n}$ where $n\geq 1$; this indexing agrees with Macaulay's inverse systems in  \cite{Northcott} and with finite sequences in \cite{N95b}. 

We can regard   finite sequences as trivial ($s=0,\ldots,0$), geometric or 'essential',  Proposition \ref{split}. 
Geometric  sequences are those of high school, defined  by $s_0\neq 0$ and a common ratio. Equivalently, they satisfy  $\LC_n=\cdots=\LC_1=1$. 'Essential' sequences on the other hand satisfy $\LC_n>\LC_1\geq 0$ and  predominate: geometric sequences may become essential on adding a  term, but never the reverse, see Proposition \ref{transition}. We summarise this  using a state diagram ('I' is the start state, 'G' denotes 'geometric' and 'E' denotes 'essential'; we have suppressed transitions between the same state):

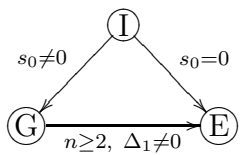
\begin{tabbing}\hspace{6cm}\=\\
\>\xymatrix{& *+[o][F-]{\rm I}\ar[dl]_{s_0\neq 0} \ar[dr]^{s_0= 0}\\
 *+[o][F-]{\rm G} \ar[rr]_{n\geq 2,\ \Delta_1\neq 0} &&*+[o][F-]{\rm E} }
\\
\end{tabbing}
where $s\neq 0,\ldots,0$ and $\Delta_1$ denotes a 'discrepancy'. Unfortunately this subdivision of sequences does  not appear in  \cite{Ma69}, which renders the Berlekamp-Massey   algorithm harder to understand. If $n\geq 2$ and $s=s_0,\ldots,s_{1-n}$ is  essential  then $$n'=\max\{1\leq j<n: \LC_j<\LC_n\}$$ is a well-defined integer, $1\leq n'<n$,  and  we have the important   subsequence
$s_0,\ldots, s_{1-n'}$.

We next discuss partial quotients  (when $S\neq 0,0,\ldots$)  as these are  classical and less detailed, being based on division in $\F((x^{-1}))$.  First we treat the  base cases in Propositions \ref{newcf0},  \ref{newcf<0}. We obtain an inductive proof of the first part of  \cite[Theorem 1]{Nied87} and its converse, Theorem \ref{simplicityitself}. This gives a similar state diagram for $S$, Proposition \ref{easyconseq}:
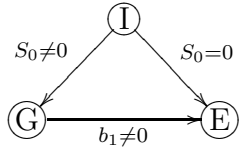
\begin{tabbing}\hspace{6cm}\=\\
\>\xymatrix{& *+[o][F-]{\rm I}\ar[dl]_{S_0\neq 0} \ar[dr]^{S_0= 0}\\
 *+[o][F-]{\rm G}  \ar[rr]_{b_1\neq 0} &&*+[o][F-]{\rm E}}
\end{tabbing}

Then we revisit \cite{N95b}, restricting to geometric and essential sequences over $\D$ only.   This new approach  is simpler, see Theorem \ref{indthm}; the corresponding Algorithm \ref{calPA} is valid for all sequences and is virtually identical to \cite[Algorithm 4.6]{N95b}; we compute $\mu=(\mu_1,\mu_2)$ rather than $(\mu_1,x\mu_2)$. (Apart from Lemmas \ref{s_0=1} and \ref{template} --- which the interested reader may verify ---  this paper is independent of \cite{N95b}.)

For sequences over a field, there is a 'normalised  Algorithm \ref{calPA}' which computes a monic $\mu_1$, Proposition \ref{monic}. This   has  been implemented  in COCOA, \cite{COCOA}. We also prove an identity for the sum of the linear complexities of $s$. This  seems to be new and gives  a simple analysis of  Algorithm \ref{calPA}, see Proposition \ref{stable}.  \\

We use Algorithm \ref{calPA} to define an element  $\nabla_s\in\D\setminus\{0\}$ and  prove the identity
\begin{eqnarray}\label{nablaeqn}\mu_2\,\mu'_1-\mu_1\,\mu'_2=\nabla_s
\end{eqnarray}
 where  $\mu'$ is either $(1,0)$ or  a minimal solution for $s_0,\ldots,s_{1-n'}$. This is our analogue of Identity (\ref{cfid}) and the identity of Berlekamp mentioned above. 
Identity (\ref{nablaeqn}) easily implies that for any  $f=(f_1,f_2)\in\D[x]^{\,2}$
\begin{eqnarray}\label{basic}\nabla_s\, f_1=m'\,\mu_1- m\,\mu'_1
\end{eqnarray} 
where $m=f_2\,\mu_1-f_1\,\mu_2$ and $m'=f_2\,\mu'_1-f_1\,\mu'_2$. This is a special case of a pairing $\D[x]^{\,2}\times\D[x]^{\,2}\ra\D[x]$ defined by $\mu$ and $\mu'$. 

In fact, essential sequences  exhibit a 'minimal system', a stronger property than (\ref{basic}),  Definition \ref{minsys}.
 We show that if $f$ is a solution and we have a minimal system, then these multipliers  (i) satisfy degree bounds  and (ii)  are unique when the degree of $f_1$ is at most $n$; in this case we call (\ref{basic}) a  'decomposition' of $\nabla_s\,f_1$. And $\nabla\,f_2$ satisfies a similar identity with the same multipliers i.e. we have a decomposition of  $\nabla_s\,f$. This  yields the required description of all solutions when we have a minimal system, see Corollary \ref{mpsol}. Partial quotients also exhibit a minimal system. In this way, we generalise the second part of  \cite[Theorem 1]{Nied87}. It also  strengthens \cite[Theorem 4.17]{N95b} and has a simpler proof.

There are lacunae for geometric sequences as they  do not have a minimal system. However this does not embarrass us, as  using $(q_1^{(0)},q_2^{(0)})=(\mu_1',\mu_2')=(1,0)$ enables us to give alternative proofs in both the partial quotient  and finite sequence contexts.   Secondly, over a domain $\D$, we  have to work with 'pseudo-geometric' sequences as the leading coefficient of $\mu_1$ may not be a unit of $\D$. As these sequences are inherently simpler and easier to treat than the essential ones, we always discuss them first.\\

We have  included some applications. We show that for a sequence $s$ over a field  and  solution $(f_1,f_2)$ such that the degree of $f_1$ is at most $n-\LC_n$ (i)  $\gcd(f_1,f_2)=1$ implies that $f$ is a minimal solution and (ii)  the multipliers of Identity (\ref{basic}) satisfy  $\gcd(m,m')=\gcd(f_1,f_2)$.  We apply (i) to linear recurring sequences. We relate our  minimal polynomials and partial quotients, Corollary \ref{lrsappl}. We also give a precise count of the number of solutions when $|\F|<\infty$.

For our final application, we revisit some work of Salagean, \cite{Salagean}. 
Let $a\in \D$ be arbitrary and suppose that $\mu_1(a)=0$.  We show that Identity (\ref{basic}) implies that the lower bound for the degree of an annihilating polynomial of $s$ which does not vanish at $a$ is $M=\LC_n+\max\{n+1-2\LC_n,0\}$. We exhibit a solution of minimal degree
$$x^M\,\mu-\mu'$$
--- the polynomial $x^M\,\mu_1-\mu'_1$ does not vanish at $a$ by Identity (\ref{basic}).
 Algorithm \ref{calPA+} is a one-line extension of Algorithm \ref{calPA} and is simpler than \cite[Algorithm 3.2]{Salagean}. We also derive  the corresponding numerator. In fact the bound in Theorem \ref{myversion}  and the set of minimal polynomials in Corollary \ref{factorialchar}  were stated without proof  in \cite{Salagean} and used to justify Algorithm 3.2,  {\em loc. cit.}\\

We thank the anonymous referee for  a number of useful comments and suggestions which improved the presentation, and also the members of {\em Projet Secret} at INRIA, Rocquencourt for their  hospitality.

 \subsection{Standard Notation}
 For any set $E$ containing 0, $E^\times=E\setminus\{0\}$ so that $\N^\times=\{1,2\ldots\}$. As usual, $\sum_\emptyset=0$.
 
   Throughout the paper, $\D$ is a  commutative domain with $1\neq 0$ and $\R=\D[x]$. For any $a\in\D^\times$ and $A\subseteq \D$, $a\,A=\{a\,b :b\in A\}$. For $f\in\R$, $|f|$ is the {\em degree} of $f\in \R$, with $|0|=-\infty$; the usual rules for arithmetic involving $-\infty$ apply.  If $f\in\R^\times$, $\mathrm{lc}(f)$ is the {\em leading coefficient} of $f$.   We  often write  $f=x^kg+h$ if $f(x)=x^kg(x)+h(x)$ where  $k\in\N$ and $g,h\in \R$. For $f,g\in\R$, their product is written $f\,g$ and we regard $\R^2$ as an $\R$-module via $f(g,h)=(f\,g,f\,h)$. 
 
 A non-zero {\em formal negative Laurent series} over $\D$ is   $L=\sum_{-\infty<i\leq k} L_i\,x^i\in \D((x^{-1}))$ where $k\in\,\Z$, $L_i\in \D$ and $L_k\neq 0$; we write   $$\vv(L)=k \ \ \ \mbox{   and   }\ \ \ [L]=\sum_{i=1}^k L_i\,x^i\in x\,\R$$ 
 i.e.  $\vv:\D((x^{-1}))\ra \{-\infty\}\cup\,\Z$ is the 
 {\em exponential valuation}, with  $\vv(L)=-\infty\Leftrightarrow L=0$; $\vv$ coincides with  $|\ \, |$ on $\R$. 
It is elementary that $\vv(L\cdot M)=\vv(L)+\vv(M)$, $\vv(L+M)\leq \max\{\vv(L),\vv(M)\}$ and $\vv(L+M)= \max\{\vv(L),\vv(M)\}$ if $\vv(L)\neq \vv(M)$. We also write $\vv$ for the restriction of $\vv$ to $\D[x^{-1}]\subset \D((x^{-1}))$. 
 We regard $\D((x^{-1}))$ as $\D[[x^{-1}]]+ x\,\R$ and use $\cdot$ for multiplication in $\D((x^{-1}))$.
 
We denote an arbitrary field by $\F$. 
For  continued fractions in $\F[[x^{-1}]]$ we use  $\Pol(L)=\sum_{i=0}^{\vv(L)} L_i\,x^i\in\F[x]$. As usual, $\F(x)\subset \F((x^{-1}))$ is the subfield of  rational functions over $\F$.
\subsection{Guide to Additional Notation}
We include a  table of additional symbols used in the paper to aid the reader.

\begin{center}
\begin{tabular}{|c|l|}\hline
Symbol & Meaning \\\hline\hline
$0^n$ & sequence of $n$ zeroes\\\hline
$a,b,c$ & elements of $\D$\\\hline
$a_{i}$ & $[a_1,a_2,\ldots]$ is the continued fraction expansion of $\ul{S}$ and $i\geq 1$\\\hline
$\Ann(s)$ & set of annihilating polynomials of $s$\\\hline
$b_{i}$ & Laurent series  in $i^{\mathrm{th}}$ iteration of partial quotient algorithm, $i\geq 0$\\\hline
$e,e_s$ & $n+1-2\LC_n$\\\hline
$f,g$ & elements $(f_1,f_2),\,(g_1,g_2)$ of $\R^2$\\\hline
$f_2$ & $[f_1\cdot\ul{s}]/x$\\\hline
$\langle f,g\rangle$ & $f_2\,g_1-f_1\,g_2\in\R$\,, the pairing of $f$ and $g$\\\hline
$\Id_S$ & ideal of characteristic polynomials of $S$\\\hline
$\LC,\LC(s),\LC_n$ & linear complexity of $s=s_0,\ldots,s_{1-n}$\\\hline
$\LC',\LC(s'),\LC_{n'}$ & linear complexity of $s'$, $s$ essential \\\hline
$m\,,m'$&$\langle f,\mu\rangle$, $\langle f,\mu'\rangle$ respectively\\\hline
$\Min(s)$ & set of minimal polynomials of $s$\\\hline
$n$ & a strictly positive  integer\\\hline
$n'$ & the strictly positive  integer $\max_{1\leq j<n}\{j:\LC_j<\LC_n\}$, $s$ essential\\\hline
$n_i$ & strictly positive integer $|q_1^{(i-1)}|+|q_1^{(i)}|$ or $\infty$\\\hline
$\nabla$ & $\nabla_s$ (non-zero product of discrepancies)  or $(-1)^{i-1}$\\\hline
$q^{(i)}$ & $i^{\mathrm{th}}$ partial quotient $(q_1^{(i)},q_2^{(i)}),\, i\geq 0$\\\hline
$s$ & finite sequence $s_0,\ldots,s_{1-n}$ over $\D$\\\hline
$s,s_{-n}$ & $s_0,\ldots,s_{1-n},s_{-n}$\\\hline
$\ul{s}$ & $s_0+s_{-1}x^{-1}+\cdots +s_{1-n}x^{1-n}$\\\hline
$s'$ &  the subsequence $s_0,\ldots,s_{1-n'}$ of essential $s$\\\hline
$S$ & infinite sequence over $\F$ \\\hline
$\ul{S}$ & $S_0+S_{-1}x^{-1}+\cdots $\\\hline
$S|n$ & $S_0,\ldots,S_{1-n}$ \\\hline
$T_{s}$ & triple $(n,\mu_1,\Delta_1)$ for $s$\\\hline
$T_{s}'$ & triple $(n',\mu'_1,\Delta'_1)$ for $s'$, $s$ essential\\\hline\hline
$\Delta_1=\Delta(\mu_1;s,s_{-n})$ & next discrepancy of  $\mu_1\in\Min(s)$\\\hline
$\Delta_1'$ & next discrepancy of  $\mu_1'$\\\hline
$\lambda$ & either $\mu$ or $q^{(i)}$,  as used in Section \ref{decomp1}\\\hline
$\mu$ & minimal solution $(\mu_1,\mu_2)$ for $s$ from Theorem \ref{indthm}\\\hline
$\mu_1$ & minimal polynomial of $s$ from Theorem \ref{indthm}\\\hline
$\mu_1'$ & 1 if $s$ is pseudo-geometric, or minimal polynomial of $s'$\\\hline
$\nu$ & new minimal solution obtained from $\mu,\mu'$\\\hline
$\xi$ &  solution with $\xi_1(a)\neq0$ constructed in Section 6, $a\in\D$\\\hline
$\phi,\phi',\psi$ & elements of $\R$.\\\hline\end{tabular}
\end{center}
\section{Sequence Basics}\label{basics}
\subsection{Rational Approximation and Solutions}
Given an infinite sequence $S=S_0,S_{-1},\ldots$ over $\F$, rational approximation of the generating function of $S$  and continued fractions is classical. Consider the  following problem:  for $n\geq 1$, find a rational function $x\,f_2/f_1\in\F(x)$ with $|f_2|<|f_1|$  such that \begin{eqnarray}\label{key} (x\,f_2/f_1)_i=S_i\mbox{ for } 1-n\leq i\leq 0
\end{eqnarray}
and $|f_1|$ is minimal. 
Let $\ul{S}=\sum_{i\leq 0}S_i\,x^i\in\F[[x^{-1}]]$ be  the {\em generating function} of $S$.  We can rephrase (\ref{key}) as: find  $x\,f_2/f_1$ such that
$\vv(\ul{S}-x\, f_2/f_1)\leq -n$ and $x\,f_2=[f_1\cdot\ul{S}]$. Multiplying by $f_1$, we equivalently require $(f_1,f_2)$ such that
\begin{eqnarray}\label{betterkey}\vv(f_1\cdot\ul{S}-x\,f_2)\leq |f_1|-n\mbox{ and }x\,f_2=[f_1\cdot\ul{S}].
\end{eqnarray}
 Let   $\{q_2^{(i)}/q_1^{(i)}:\ i\geq 0\}$ be the partial quotients of $x^{-1}\ul{S}$. In \cite{Nied87} (with $S=S_1,S_2,\ldots$ and  $\ul{S}=\sum_{i\geq 1}S_i\,x^{-i}\in x^{-1}\F[[x^{-1}]]$\,) Theorem 1,  {\em loc. cit.} shows that
 
(i) if $|q_1^{(i-1)}|+|q_1^{(i)}|\leq n< |q_1^{(i)}|+|q_1^{(i+1)}|$ then $(q_1^{(i)},q_2^{(i)})$ solves (\ref{betterkey}); 

(ii)  $q_1^{(i)}$ is an '$n^{\mathrm{th}}$ minimal polynomial' of $S$, \cite[p. 39]{Nied87};

 (iii)  all  $n^{\mathrm{th}}$ minimal polynomials of $S$ can be expressed in terms of $q_1^{(i)}$ and $q_1^{(i-1)}$.

\subsection{Linear Recurring Sequences}
For an infinite sequence $S$ over  $\F$, we easily have $\ul{S}=x\,\psi/\varphi\in\F(x)$  for some $\psi\in\F[x]$ with $|\psi|<|\varphi|=d$ if and only if $(\varphi\cdot \ul{S})_i=0$ for $i\leq 0$ and $x\,\psi=[\varphi\cdot \ul{S}]$.
Now $(\varphi\cdot \ul{S})_i=\varphi_dS_{i-d}+\varphi_{d-1}S_{i-d+1}+\cdots+\varphi_0S_i$ so that $S_0,\ldots,S_{1-d}$, $\varphi$ and the equation
\begin{eqnarray}\label{lrs}
S_{i-d}=-(\varphi_{d-1}S_{i-d+1}+\cdots+\varphi_0S_i)/\varphi_d\ \ \ \mbox\,{ for }\ \ i\leq 0
\end{eqnarray}
uniquely determine all subsequent terms of $S$;  $\varphi$ is called a {\em characteristic polynomial} of the {\em linear recurring sequence} $S$. It is well-known that these polynomials form a {\em (principal) ideal} $\Id_S$ of $\F[x]$, generated by a {\em minimal polynomial} of $S$.
 
The situation is similar for $n\geq 1$ and a finite sequence $s=s_0,\ldots,s_{1-n}$ over $\F$ i.e. $s_i\in\F$. If $\varphi\in\F[x]$,  $1\leq d=|\varphi|<n$ and
$$s_{i-d}=-(\varphi_{d-1}s_{i-d+1}+\cdots+\varphi_0s_i)/\varphi_d\ \ \ \mbox\,{for }\ \ d+1-n\leq i\leq 0
$$
then $\varphi$ and $s_0,\ldots,s_{1-d}$ uniquely determine $s_{-d},\ldots,s_{1-n}$ and $\varphi$ is often called a 'characteristic polynomial' of $s$. However, these do not form an ideal of $\F[x]$; the reader may easily find  examples for which a sum of characteristic polynomials  of $s$ is not a characteristic polynomial.

\subsection{Annihilating Polynomials and Solutions}
Let  $n\geq 1$ and $s=s_0,\ldots,s_{1-n}$ be a finite sequence over $\D$ i.e.  $s_i\in\D$; $s$ is {\em trivial} if $s=0,\ldots,0=0^n$. The {\em generating function} of $s$ is $\ul{s}=\sum_{i=1-n}^0s_i\,x^i\in\D[x^{-1}]$. We put $\vv=\vv(\ul{s})$ if $s$ is understood, so that  if $s$ is non-trivial then $1-n\leq \vv\leq 0$. 
\begin{definition}(\cite[Definition 2.7]{N95b}) \label{anndefn} Let $\varphi\in \R$, $d=|\varphi|$ and   $\varphi=\sum_{i=0}^d\varphi_i\,x^i$. If $n\geq 1$ and  $s=s_0,\ldots,s_{1-n}$ then $\varphi$  is an  {\em annihilating polynomial}  of $s$, written $\varphi\in\Ann(s)$,  if 
\begin{eqnarray}\label{anncond}(\varphi\cdot \ul{s})_i= \varphi_ds_{i-d}+\varphi_{d-1}s_{i-d+1}+\cdots +\varphi_0s_i=0\ \ \ \mbox{ for }d+1-n\leq i\leq 0.
\end{eqnarray} 

\end{definition}
We note that  (\ref{anncond}) is vacuously satisfied if $d=|\varphi|\geq n$, so that the previous definition is equivalent to \cite[Definition 2.7]{N95b} and $$
\{\varphi\in\R^\times:|\varphi|\geq n\}\subseteq \Ann(s)^\times=\{\varphi\in \R^\times:  (\varphi\cdot \ul{s})_i=0\mbox{ for } |\varphi|+1-n\leq i\leq 0\}.$$ 
 
 We prefer  'annihilating polynomial' to 'characteristic polynomial' as we do not insist that $\lc(\varphi)$ be  a unit of $\D$. Further, we may be unable to express $s_{1-n}$ as a linear combination of $s_0,\ldots,s_{2-n}$; e.g. if $s=0^{n-1},1$ then $s_{1-n}$ is not a linear combination of 0's; if $\D=\Z$ and $s=2,1$ we cannot express $1$ as a multiple of $2$ in $\D$.

As in (\ref{betterkey})  we now have:
\bp \label{solchar} For $n\geq 1$ and a sequence $s=s_0,\ldots,s_{1-n}$ over $\D$\begin{eqnarray*}f_1\in\Ann(s)^\times\mbox{  if and only if } f_1\in\R^\times,  \vv(f_1\cdot\ul{s}-x\,f_2)\leq |f_1|-n \mbox{ and }x\,f_2=[f_1\cdot\ul{s}].
\end{eqnarray*}
We say that $f=(f_1,f_2)\in\R^\times\times\R$ (or $x\,f_2/f_1$ if $f_1^{-1}$ exists) is a {\em solution for $s$}. 
\ep
If $(\varphi,\psi)\in\R^2$ is a solution for $s$ and $d=|\varphi|\geq 0$ then $|\psi|=\vv+d-1$. Also $\varphi$ and $s_0,\ldots,s_{1-d}$ determine $\psi$, since for $1\leq i\leq \vv+d$ we have $\psi_{i-1}=(x\,\psi)_i=[\varphi\cdot\ul{s}]_i=\sum_{j=0}^d\varphi_j\,s_{i-j}$ where $1-d\leq i-j\leq 0$. We include a proof of the following for completeness.
\bp Let $(\varphi,\psi)$ be a solution for $s=s_0,\ldots,s_{1-n}$.  If  $1\leq d=|\varphi|<n$ and $\lc(\varphi)$ is a unit of $\D$ then (i) $\varphi^{-1}\in\D[[x^{-1}]]$;  (ii) $\varphi$, $s_0,\ldots,s_{1-d}$ determine $s_{-d},\ldots,s_{1-n}$. 
\ep
\bpr  If $\sigma=1-\varphi\, x^{-d}/\varphi_d \in\D[x^{-1}]$ then $\varphi=\varphi_d\,x^d(1-\sigma)$, $\sigma\neq 1$ since $\varphi\neq 0$ and  $$\frac{1}{\varphi}=\frac{1}{\varphi_d\,x^d(1-\sigma)}=\varphi_d^{-1}\,x^{-d}(1+\sigma+\sigma^2+\cdots)\in\D[[x^{-1}]].$$ Thus $\vv(\ul{s}-x\,\psi/\varphi)=\vv(\varphi\cdot\ul{s}-x\,\psi)+\vv(1/\varphi)\leq |\varphi|-n-\vv(\varphi)=-n$, which implies that $(x\,\psi/\varphi)_i=s_i$ for $n-1\leq i\leq 0$. We know that $\varphi$ and $s_0,\ldots,s_{1-d}$ determine $\psi$. Hence $\varphi$ and $s_0,\ldots,s_{1-d}$  determine  $s_{-d},\ldots,s_{1-n}$.
\epr
 Thus  if $f$ is a solution, $|f_1|<n$ and $\lc(f_1)$ is a unit of $\D$ then   $s_0,\ldots,s_{1-n}$ and $f_1$ define  a linear recurring sequence $S_f$ with $\ul{S}_f=x\,f_2/f_1$ and $f_1\in\Id_{S_f}$.
\subsection{First Examples and the Key  Lemma \ref{onemore2}} Our first examples include the inductive bases for various results below. 
\be \label{first} (i) The (finite) geometric sequence of length $n$ with  common multiple $m\in\D^\times$ is given by  $s_0=1$ and  $s_i=m^{-i}$ for $1-n\leq i\leq -1$. It will be convenient to allow $m=0$ as well. Then $\vv=0$ and $\ul{s}=m^{n-1}x^{1-n}+\cdots +1$. If $|x-m|+1-n=2-n\leq i\leq 0$ then $((x-m)\cdot\ul{s})_i=(x\cdot\ul{s})_i-(m\cdot\ul{s})_i=s_{i-1}-ms_i=0$, so that $x-m\in\Ann(s)^\times$. As $x-m$ is invertible, the corresponding solution is $x/(x-m)$.

(ii) Let $n\geq 2$, $a\in\D^\times$ and $s=0^{n-1},a$. We have $\ul{s}=ax^{1-n}$, $x^n=x^{1-\vv}\in\Ann(s)^\times$ and   $ax/x^n$ is a solution for $s$, but cannot express $s_{1-n}$ as a linear combination of zeroes.

(iii) Let  $\D=\Z$ and $s=2,1$. We have the solution $(2x-1,2)$ but  cannot express $s_{-1}$ in terms of $s_0$. 
\eex

For $s=s_0,\ldots,s_{1-n}$ and arbitrary $a\in \D$,  $t=s,a$ is the sequence $s_0,\ldots,s_{1-n},a$. Given a solution $g$ for $s$ we want to construct a solution $h$ for $t$. 
We begin with first components. 
 It is clear that $\Ann(t)^\times\subseteq \Ann(s)^\times$. Suppose that $g_1\in\Ann(s)^\times$. Then $g_1\in\Ann(t)$ if and only if $(g_1\cdot\ul{t})_i=0$ for $|g_1|-n\leq i\leq 0$. If $|g_1|+1-n\leq i\leq 0$ and $0\leq j\leq |g_1|$ then 
 $1-n\leq i-j\leq 0$ and so $t_{i-j}=s_{i-j}$. Hence $(g_1\cdot\ul{t})_i=(g_1\cdot\ul{s})_i=0$ for $|g_1|+1-n\leq i\leq 0$ and $g_1\in\Ann(t)$ if and  only if $(g_1\cdot\ul{t})_{|g_1|-n}=0$.  
\bd (\cite[Definition 2.10]{N95b}, cf. \cite{Ma69}).
Let $n\geq 1$,   $s=s_0,\ldots,s_{1-n}$ and $g_1\in \R^\times$. For  arbitrary $a\in\D$ and $t=s,a$ the {\em discrepancy} of $g_1$ and $t$ is  $\Delta(g_1;t)=(g_1\cdot\ul{t})_{|g_1|-n}$. 
\ed 
In general, if $g_1\in\Ann(s)$ then  $g_1\cdot\ul{t}=G+\Delta(g_1;t)\,x^{|g_1|-n}+[g_1\cdot\ul{t}]$ where $\vv(G)< |g_1|-n$. 
We recall the proof of the following lemma from \cite{N99b} as it shows the usefulness of second components. Also, the polynomial $g_2\,h_1-g_1\,h_2$ of the proof will reappear later.
\bl\label{onemore2}(\cite[Lemma 5.2]{N99b}, cf. \cite[Theorem 1]{Ma69})   Let   
$n\geq 1$, $s=s_0\ldots,s_{1-n}$ and $g_1\in\Ann(s)^\times$. If $t=s,a$ and $g_1\not\in\Ann(t)$ then for any $h_1\in\Ann(t)^\times$ we have $|h_1|\geq n+1-|g_1|$. 
\el
\bpr  Let  $\Delta=\Delta(g_1;t)\neq 0$.  We have $g_1\cdot\ul{t}=G+\Delta\, x^{|g_1|-n}+x\,g_2$ where $\vv(G)<|g_1|-n$ and $x\,g_2=[g_1\cdot\ul{t}]$. Also $h_1\cdot\ul{t}=H+x\,h_2$ where $\vv(H)<|h_1|-n$ and $x\,h_2=[h_1\cdot\ul{t}]$.   
Put $\varphi=g_2\,h_1-g_1\,h_2\in \R$. Then
\begin{eqnarray*}x\,\varphi&=&(x\,g_2)\,h_1-g_1\,(x\,h_2)\\
&=&(g_1\cdot\ul{t}-G-\Delta\, x^{|g_1|-n})h_1-g_1(h_1\cdot\ul{t}-H)
= -G\,h_1+g_1H-\Delta\, x^{|g_1|-n}h_1
\end{eqnarray*}
where $\vv(-Gh_1+g_1H)<|g_1|-n+|h_1|$ and $x\,\varphi\neq 0$ since $\D$ has no zero divisors. Hence $|g_1|+|h_1|-n=|x\,\varphi|\geq 1$.
\epr
For Example \ref{first}(i), if $h$ is a solution for $s$ then $|h_1|\geq 1$ since $s$ is non-trivial.
For Example \ref{first}(ii) with $n\geq 2$,   $(1,0)$ is a solution for $0,\ldots,0$ but not for $s$ since $a\neq 0$, so if $h$ is a solution for $s$  then $|h_1|\geq -\vv+1=n$. This can also be proved directly, see  \cite[Proposition 3.5(c)]{N95b}.\\

\subsection{Linear Complexity and Minimal Solutions}
Next we discuss minimality. Firstly,  $\Ann(s)^\times\neq \emptyset$ since any polynomial of degree $n$ annihilates $s$ and the following definition makes sense.
\begin{definition}\label{mindefn} (\cite[Definition 3.1]{N95b})  Let $n\geq 1$ and $s=s_0\ldots,s_{1-n}$. The {\em linear complexity} $s$ is 
$$\LC_n=\LC(s)=\min\{|f_1|:\ f_1\in \Ann(s)^\times\}.$$ 
We say that $f_1$ is a {\em minimal polynomial (MP)} of $s$ if $f_1\in \Ann(s)^\times$ and $|f_1|=\LC(s)$. We write $\Min(s)$ for the {\em set of minimal polynomials of $s$} and say that  $f\in\R^2$ is a {\em minimal solution for $s$} if it is a solution for $s$ and $f_1\in\Min(s)$.
\end{definition}
 It is important to note that linear complexity and minimality are defined independently of how  solutions are obtained. Of course, $0/1$ is a minimal solution  for any sequence of zeroes, and $\LC(s)=0$ if and only if $s$ is trivial.  
For Example \ref{first}(ii), $x^n\in\Ann(s)$, so $\LC_n=n$ by Lemma \ref{onemore2}. 
The function $\LC$ is a non-decreasing function of $n$ and $\LC_n\leq n$. If $h_1\in\Ann(s,a)$ and $|h_1|=\LC(s)$ then $h_1\in\Min(s,a)$ since $\LC(s,a)\geq \LC(s)=|h_1|\geq \LC(s,a)$.\\

 We repeat the proof of the next result from \cite{N95b} for the convenience of the reader.
\bp \label{minfield} (\cite[Corollary 3.24]{N95b}) Let $n\geq 1$, $s=s_0,\ldots,s_{1-n}$ be a sequence over $\F$, $f$ be a solution for $s$ and $d=\gcd(f_1,f_2)$. Then (i) $f/d$ is a solution for $s$; (ii) if $f$ is a minimal solution for $s$  then $|f_1|\leq n$ and $d=1$.
\ep
\bpr  Let   $f_1\cdot\ul{s}=F+xf_2$ where $\vv(F)\leq |f_1|-n$ and $g=f/d$. Then  $F/d\in\F[[x^{-1}]]$ and $g_1\cdot\ul{s}=F/d+xg_2$ where $\vv(F/d)=\vv(F)-|d|\leq |f_1|-n-|d|=|g_1|-n$. Hence $g_1\in\Ann(s)$ and $[g_1\cdot\ul{s}]=xg_2$ i.e. $g$ is a solution for $s$.
 (ii) If $f$ is a minimal solution then $|f_1|=\LC_n\leq n$ and $|f_1|=\LC_n\leq|g_1|= |f_1|-|d|$ and hence $|d|=0$.
\epr
So if $s$ is a sequence over $\F$, $f$ is any minimal solution for $s$ and $\ul{S}_f=xf_2/f_1$,  $\Id_{S_f}=f_1\,\F[x]$ i.e. $f_1$ is a minimal polynomial for $S_f$.  This justifies our use of the term 'minimal polynomial' of $s$. The converse of Proposition \ref{minfield}(ii) fails:
 let $n\geq 2$ and $s_i=m^{-i}$ for $1-n\leq i\leq 0$ where $m\in\F^\times$; $x-m\in\Min(s)$ and $\LC_n=1$. Now $f_1=x^n\in\Ann(s)$, $f_2=m^0x^{n-1}+\cdots +m^{n-1}$ and  $\gcd(f_1,f_2)=1$ since $m^{n-1}\neq 0$, but $f_1\not\in\Min(s)$.

We will need to single out two kinds of non-trivial sequences: 
\bd  Let  $n\geq 1$ and $s=s_0,\ldots,s_{1-n}$ be a sequence over $\D$. We call $s$ {\em pseudo-geometric} if $\LC_n=\cdots=\LC_1=1$, and  {\em essential} if $n\geq 2$ and $\LC_n>\LC_1\geq 0$.  
 \ed
 
 Any geometric sequence $s$ is pseudo-geometric since $x-s_{-1}/s_0\in\Min(s)$. A non-trivial sequence $s=a,0^{n-1}$  is   pseudo-geometric. In general, $s$ is  pseudo-geometric if and only if   $s_0x-s_{-1}\in\Min(s)$.  Conversely, if $n\geq 2$, $s_0$ is a unit of $\D$ and $s$ is pseudo-geometric  then $s$ is a geometric sequence with  common ratio $s_{-1}/s_0$.  
 
Essential sequences were motivated by the need for the integer  $$n'=\max_{1\leq j<n}\{j: \LC_j<\LC_n\}$$ 
to be well-defined; now $s'=s_0,\ldots,s_{1-n'}$ is a well-defined, proper  subsequence of $s$ as $1\leq n'<n$. 

{\bf N.B.} {\em For $n\geq 2$, the sequence  $s=0^{n-1},a$ of Example \ref{first}(ii) is essential since $\LC_n=n>\LC_1=0$,  $1\leq n'=n-1<n$ (and moreover $s'=0^{n-1}$  has   minimal solution $0/1$).} On the other hand,  if $s$ is pseudo-geometric then $n'$ is undefined.

We can now formally state our subdivision of sequences.

 \bp \label{split} A sequence over $D$ is either trivial, pseudo-geometric or essential.
 \ep
 \bpr  We have $\LC_1=\cdots=\LC_n=0$ if and only if $s$ is trivial. Hence if $s$ non-trivial, $\LC_i>0$ for some $i$, $1\leq i\leq n$. If $\LC_1\neq 0$  then $\LC_1=1$ since $\LC_i\leq i$ for all $i$. If $1=\LC_2=\cdots=\LC_n$ then $s$ is pseudo-geometric; otherwise if $i$ is the first integer with $\LC_i>1$ then $\LC_n\geq \LC_i>1$ i.e. $s$ is essential. Finally, if $\LC_1=0$ then $\LC_i\neq 0$ for some $i$, $2\leq i\leq n$ as $s$ is non-trivial, so $\LC_n\geq \LC_i>\LC_1=0$ and $s$ is essential.
\epr
\bp\label{transition} If $n\geq 2$, $s=s_0,\ldots,s_{1-n}$ is pseudo-geometric and $\mu_1\in\Min(s)$ satisfies  $\Delta(\mu_1;s,s_{-n})\neq 0$ then $s,s_{-n}$ is essential.
\ep
\bpr  By Lemma \ref{onemore2}, $\LC_{n+1}\geq n+1-\LC_n=n\geq 2>1=\LC_1$.   \epr
 The first state diagram of the Introduction illustrates this transition on adding a term.

\section{Minimal Solutions via Partial Quotients}\label{cf}
Here we revisit the first part of \cite[Theorem 1]{Nied87}. Let $S$ be an infinite sequence over  $\F$, $\ul{S}$ its generating function  and $n\geq 1$. Our goal is to show that a certain partial quotient of $\ul{S}$ (depending on $n$) is a minimal solution for $S|n=S_0,\ldots,S_{1-n}$.  In particular, we relate $\LC(S|n)$ to the degrees of the partial quotients of $\ul{S}$.

We recall the construction of the partial quotients of $\ul{S}$,  their basic identities and properties. We work through \cite[Example 1]{WS}. Then we discuss geometric sequences and  $0^{n-1},S_{1-n}$, where $S_{1-n}\in\F^\times$ and $n\geq 2$. These form our inductive basis for the main Theorem \ref{simplicityitself}. When $S|n$ is essential, we  determine $\max_{1\leq j<n}\{\LC_j<\LC_n\}$  and prove an identity for any $f\in\R^2$.
\subsection{Continued Fractions}
We use the formulation of continued fractions in $\F[[x^{-1}]]$ from  \cite{Nied87}; in particular, we also use  $\Pol(L)=\sum_{i=0}^{\vv(L)}L_ix^i\in\R$ for  $L\in\F((x^{-1}))$.
It is well known that $\ul{S}$ has the unique continued fraction expansion $0+x/(a_1+1/(a_2+\cdots))=[0,a_1,a_2,\ldots]$ 
where, if $a_i\in\R$ exists, then $|a_i|\geq 1$.
The $a_i$ are obtained using division in the field $\F((x^{-1}))$ as follows:
\begin{tabbing}
$a_0\la0;\ b_0\la x^{-1}\ul{S}$; $q^{(-1)}\la(0,1)$; $q^{(0)}\la(1,0)$;\\\\

$i\la0$;\\
{\tt while} \= $(b_i\neq 0)$\\\\ 
\>$\lceil$\ \= $a_{i+1}\la\Pol (b_i^{-1})$;  $q^{(i+1)}\la a_{i+1}q^{(i)}+q^{(i-1)};$\\\\
\> \> $b_{i+1}\la b_i^{-1}-a_{i+1}$; $i\la i+1;\ \rfloor$
\end{tabbing}

The  {\em partial quotients}  of $\ul{S}$ are $\{q^{(i)}\in\R^2:\ i\geq 0\}$. Put $|q_1^{(-1)}|=0$ and if $b_i=0$ (i.e. $a_{i+1}$ and $q^{(i+1)}$ do not exist) put $|a_{i+1}|=|q_1^{(i+1)}|=\infty$.  In the following well-known result, Part (iv) on numerators is probably well-known, but does not appear in \cite{Nied87}.

\bt \label{cfbasics} If $i\geq 1$, $q=q^{(i)}$ exists and $q'=q^{(i-1)}$ then  $|q_1|=\sum_{j=1}^i|a_j|\geq 1$. In particular, if $i\geq 0$  then  $1-|q_1^{(i+1)}|\leq 0$. 

If  $i\geq 0$ and $q$ exists then 
 
(i)  
$q_2\,q_1'-q_1\,q_2'=(-1)^{i-1}$  and $\gcd(q_1,q_1')=\gcd(q_2,q_2')=1$;

(ii) $$   
\ul{S}=\frac{x\,(q_2+b_i\,q'_2)}{q_1+b_iq'_1}\,;$$

(iii)  $\vv(q_1\cdot\ul{S}-x\,q_2)=1-|q_1^{(i+1)}|$ so that $\Pol(q_1\cdot\ul{S}-x\,q_2)=0$;

(iv) $|a_1|=1-\vv$,  $|q_2|=\sum_{j=2}^i|a_j|=|q_1|-|a_1|$ if  $i\geq 1$ and $\Pol(q_1\cdot\ul{S})=[q_1\cdot\ul{S}]=x\,q_2$.\\

If $i\geq 1$ is the first index for which $b_i=0$ then  

(v) $\ul{S}=x\,q_2/q_1$ and  $\vv(q_1\cdot\ul{S}-x\,q_2)=-\infty=1-|q_1^{(i+1)}|$;

(vi) $\Pol(q_1\cdot\ul{S})=x\,q_2$ and $|q_2|=\vv+|q_1|-1<|q_1|$.
\et
\bpr  (i)-(iii) Use induction and properties of the exponential valuation as in \cite{Nied87}.  (iv) For $i\geq 1$, $|q_2|=\sum_{j=2}^i|a_j|$ is an easy induction.  We have $|a_1|=|\Pol(x/\ul{S})|=1-\vv$ so that $|q_2|=|q_1|-|a_1|=|q_1|+\vv-1$ and 
$$\Pol(q_1\cdot\ul{S})=\Pol(q_1\cdot\ul{S}-x\,q_2)+\sum_{j=0}^{\vv+|q_1|}(x\,q_2)_j\,x^j=x\sum_{k=-1}^{|q_2|}(q_2)_k\,x^k=x\,q_2$$ by (iii). 
 (v) If $i\geq 1$, $b_{i-1}\neq 0$ and $b_i=0$ then  
 $$q=q+(b_{i-1}^{-1}-a_i)q'=q-a_iq'+b_{i-1}^{-1}q'=q^{(i-2)}+b_{i-1}^{-1}q'$$ and rearranging gives $$\frac{x\,q_2}{q_1}=\frac{x\,(q'_2+b_{i-1}q^{(i-2)}_2)}{q'_1+b_{i-1}q^{(i-2)}_1}=\ul{S}$$
by (ii).  (vi) Immediate.
\epr
Next we define a partition of $\N^\times$. Let $n_0=1$ and  for $i\geq 1$, define  $n_i=n_i(S)$ by 
$$n_i=|q_1^{(i-1)}|+|q_1^{(i)}|.$$

From Theorem \ref{cfbasics} we have $1=n_0\leq n_1<n_2<\cdots$ and $\{[n_i,n_{i+1}): \ i\geq 0\}$ is a {\em partition} of $\N^\times$ (except that $[n_0,n_1)=\emptyset$ if $n_1=1$) for if $i\in\N$ is largest such that $n_i\leq n$ then $n\in[n_i,n_{i+1})$ and $i$ is clearly unique.  Thus if $\ul{S}=0$,   $n_1=\infty$ and for all $n\geq 1$,  $0/1$ is a minimal solution for $S|n$. \\
 
 The next example is \cite[Example, p. 21]{WS}. Here $\F=\F_2[x]/(x^4+x+1)$ and $\alpha$ generates $\F^\times$. For the table of $\F^\times$ as polynomials in $\alpha$, see \cite[p. 85]{McWS}.
\begin{example}\label{WSexample1} Let  $\ul{S}=x({\alpha^5x^2+\alpha^2x+\alpha^{10}})/({x^3+\alpha^6x^2+\alpha^3x+\alpha^{13}})\in\F(x)$. Direct calculation  gives 
\begin{center}
\begin{tabular}{|l|l|l|l|}\hline
$i$  & $b_i$&$a_{i+1}$  &$b_{i+1}=b_i^{-1}-a_{i+1}$\\\hline\hline
$0$ &$x^{-1}\ul{S}$  &$\alpha^{10}x+\alpha^{14}$ &$(\alpha^6 x+\alpha^{10})/(\alpha^5x^2+\alpha^{2}x+\alpha^{10})$\\\hline
$1$ &$(\alpha^6 x+\alpha^{10})/(\alpha^5x^2+\alpha^{2}x+\alpha^{10})$ & $\alpha^{14}x+\alpha^5$ &$\alpha^{5}/(\alpha^6x+\alpha^{10})$\\\hline
$2$ &$\alpha^{5}/(\alpha^6x+\alpha^{10})$&$ \alpha x+\alpha^5$& $0$ \\\hline
\end{tabular}
\end{center}
\begin{center}
\begin{tabular}{|l|l|l|l|}\hline
$i$ &$q^{(i-1)}$  &$q^{(i)}$   & $q^{(i+1)}=a_{i+1}\,q^{(i)}+q^{(i-1)}$\\\hline\hline
$0$ &$(0,1)$  &$(1,0)$   &$(\alpha^{10}x+\alpha^{14},1)$\\\hline
$1$ &$(1,0)$ &$(\alpha^{10}x+\alpha^{14},1)$&$(\alpha^9x^2+\alpha^{6}x+\alpha,\alpha^{14}x+\alpha^{5})$\\\hline
\end{tabular}
\end{center}
and $q^{(3)}$ is $$(\alpha^{10}x^3+\alpha^{13}x^2+\alpha^3x+\alpha^{8},x^2+\alpha^{12}x+\alpha^{5})=\alpha^{10}(x^3+\alpha^6x^2+\alpha^3x+\alpha^{13},\alpha^5x^2+\alpha^2x+\alpha^{10}).$$ Clearing denominators, this is  the extended Euclidean algorithm: $a_{i+1}$ is the quotient and $b_{i+1}$ is the remainder. We have $|q_1^{(0)}|=0$, $|q_1^{(1)}|=1$, $|q_1^{(2)}|=2$ and $|q_1^{(3)}|=3$ and $|q_1^{(4)}|=\infty$, so that the partition of $\N^\times$ defined by $S$ is $[1,3),[3,5),[5,\infty)$.
\eex
 
Our inductive proof of the first part of \cite[Theorem 1]{Nied87}  depends on  characterising solutions for $S|n$ in terms of solutions for $S$, and is proved using  Proposition \ref{solchar}.

\bl \label{Ann(S|n)} Let $\D=\F$ be a field. If $f_1\in\R^\times$, $|f_1|\leq n$ and $s=S|n$ then $[f_1\cdot\ul{S}]=[f_1\cdot\ul{s}]=x\,f_2$ say. Further $f$ is a solution for $s$ if and only if $\vv(f_1\cdot\ul{S}-x\,f_2)\leq |f_1|-n$.
\el
\bpr We have $\vv(\ul{S}-\ul{s})\leq -n$ and $[f_1\cdot(\ul{S}-\ul{s})]=0$, so $[f_1\cdot\ul{S}]=[f_1\cdot(\ul{s}+(\ul{S}-\ul{s}))]=[f_1\cdot\ul{s}]+[f_1\cdot(\ul{S}-\ul{s})]=[f_1\cdot\ul{s}]$ and  
\begin{eqnarray*}
f_1\cdot\ul{S}-x[f_1\cdot\ul{S}]&=&f_1\cdot(\ul{s}+(\ul{S}-\ul{s}))-x[f_1\cdot\ul{s}]
=f_1\cdot\ul{s}-x[f_1\cdot\ul{s}]+f_1\cdot(\ul{S}-\ul{s}).\end{eqnarray*}
Hence 
$\vv(f_1\cdot\ul{S}-x[f_1\cdot\ul{S}])\leq \max\{\vv(f_1\cdot\ul{s}-x[f_1\cdot\ul{s}]), |f_1|-n\}$ and if  $f_1\in\Ann(s)$ then $\vv(f_1\cdot\ul{S}-[f_1\cdot\ul{S}])\leq|f_1|-n$ by Proposition \ref{solchar}. The converse is proved similarly, for $f_1\cdot\ul{s}-x[f_1\cdot\ul{s}]=f_1\cdot\ul{S}-x[f_1\cdot\ul{S}]-f_1\cdot(\ul{S}-\ul{s})$.
\epr

 \subsection{Geometric Sequences}
If $S_0\neq 0$,  Theorem \ref{cfbasics} implies that $n_1=|a_1|=1-\vv=1$.
\bp\label{newcf0} Let $S$ be an infinite sequence over a field $\F$ such that $S_0\neq 0$, $q=q^{(1)}$ and $q'=(1,0)$. The following are equivalent 

(i)  $n\in[n_1,n_2)$; 

(ii) $q'$ is not a solution for $s=S|n$ and $q$ is a minimal solution for $s$.
\ep
\bpr (i) $\Rightarrow$ (ii). Firstly $q'$ is not a solution for $s$ since $S_0\neq 0$. Secondly, $|q_1|\leq n$ and 
 $$\vv(q_1\cdot \ul{S}-x\,q_2)=1-|q_1^{(2)}|=1-(n_2-1)=2-n_2\leq 1-n.$$
By Lemma \ref{Ann(S|n)}, $q$ is a solution for $s$. As $|q_1|=1$, it is a  minimal solution.  

(ii) $\Rightarrow$ (i). If $n\not\in[n_1,n_2)$ then $n\in[n_2,n_3)$  so $\vv(q_1\cdot \ul{S}-x\,q_2)=1-|q_1^{(2)}|=2-n_2>1-n$ and $q_1\not\in\Ann(s)$. In particular, $q_1\not\in\Min(s)$.
\epr 
We can say more.
\bp\label{cf0} Let $S_0\neq 0$ and $r=S_{-1}/S_0$. Then 
 $a_1=(x-r)/S_0$ and $n_2\geq 3$.
\ep
\bpr   Write $\ul{T}=\ul{S}/S_0=1+rx^{-1}+\cdots$\ ; if this is a geometric series then $\ul{T}=x/(x-r)$, $b_0=x^{-1}\ul{S}=S_0/(x-r)$,  $a_1=\Pol(b_0^{-1})=(x-r)/S_0$ and $n_2=\infty$. 

Otherwise we have $\ul{T}=1+rx^{-1}+r^2x^{-2}+\cdots+r^{n_2-2}x^{2-n_2}+T_{1-n_2}+\cdots$ where  $T_{1-n_2}\neq rT_{2-n_2}$ and $2-n_2\leq -1$.
 Now $\ul{T}=x/(x-r)+\ul{U}$ for some $\ul{U}$ with   $\vv(\ul{U})\leq 1-n_2\leq -2$ and
 $$b_0/S_0=x^{-1}\ul{S}/S_0=x^{-1}\,\ul{T}=
(x-r)^{-1}+x^{-1}\ul{U}= \frac{1+(x-r)x^{-1}\ul{U}}{x-r}=\frac{1-\ul{V}}{x-r}$$
say, where $\vv(\ul{V})\leq -2$. Thus for all $i\geq 1$ we have $\vv(\ul{V}^i)\leq -2i\leq -2$,
 $$b_0^{-1}=S_0^{-1}(x-r)/(1-\ul{V})=S_0^{-1}(x-r)(1+\ul{V}+\ul{V}^2+\cdots)$$  and $a_1=\Pol(b_0^{-1})=(x-r)/S_0$. 
 \epr
\subsection{Essential Sequences and the General Case}
If  $S_0=0$, Theorem \ref{cfbasics} implies that  $n_1=|a_1|=1-\vv>1$. 
\bp\label{newcf<0} Let $S$ be an infinite sequence over a field $\F$ such that $\ul{S}\neq 0$, $S_0=0$, $q=q^{(1)}$ and $q'=(1,0)$. The following are equivalent 

(i)  $n\in[n_1,n_2)$; 

(ii) $q'$ is not a solution for $s=S|n$ and $q$ is a minimal solution for $s$.
\ep
\bpr (i) $\Rightarrow$ (ii).  Let $n\in[n_1,n_2)$.  We have $|q_1|=1-\vv=n_1\leq n$ and 
 $$\vv(q_1\cdot\ul{S}-x\,q_2)=1-|q_1^{(2)}|=1+|q_1|-n_2\leq |q_1|-n$$
 from Theorem \ref{cfbasics}, 
  so that $q$ is a solution for $s$ by Lemma \ref{Ann(S|n)}.  As $n_1-1\geq 1$ and $1\in\Ann(S|n_1-1)\setminus\Ann(S|n_1)$,  Lemma \ref{onemore2} implies that $1-\vv=|q_1|\geq \LC_n\geq\LC_{n_1}\geq n_1-\LC_{n_1-1}=n_1=1-\vv$ so $\LC_n=1-\vv$ and $q$ is a minimal solution for $s$.

(ii) $\Rightarrow$ (i). If $n\not\in[n_1,n_2)$ then either (a) $n\in[1,n_1)$, 
 $\vv(q_1'\cdot\ul{S}-xq'_2)=1-|q_1|=1-n_1\leq -n$ and  $q'_1=1\in\Ann(s)$ or (b) 
 $n\in[n_2,n_3)$ and as in the proof of Proposition \ref{newcf0},
 $\vv(q_1\cdot \ul{S}-x\,q_2)=1-|q_1^{(2)}|=2-n_2>1-n$, so $q_1\not\in\Ann(s)$ and   $q_1\not\in\Min(s)$.
 \epr

We have now treated the case $n\in[n_1,n_2)$. Now for the general case. 
\bt\label{simplicityitself} (Cf.  \cite[Theorem 1]{Nied87},  \cite[Theorem 4]{Cheng})  Let $\ul{S}\neq0$, $i\geq 1$ and $q=q^{(i)}$,  $q'=q^{(i-1)}$. The following are equivalent:

(i)  $n\in[n_i,n_{i+1})$;

(ii) $q'$ is not a solution for $s=S|n$ and  $q$ is a minimal solution for $s$.
\et
\bpr  For $i=1$ the result follows from Propositions \ref{newcf0} and  \ref{newcf<0}. Suppose inductively that $i\geq 2$ and that the result is true for $i-1$. 

(i) $\Rightarrow$ (ii).  Let $n\in[n_i,n_{i+1})$. Then $0\leq |q_1'|=n_{i}-|q_1|\leq  n-|q_1|$ i.e. $|q_1|\leq n$ and Lemma \ref{Ann(S|n)} applies.  If  $n< n_{i+1}$ then by Theorem \ref{cfbasics}  
$$\vv(q_1\cdot\ul{S}-x\,q_2)=1-|q_1^{(i+1)}|=1-n_{i+1}+|q_1|\leq |q_1|-n$$
so $q_1\in\Ann(s)$. Likewise $|q_1'|<|q_1|\leq n$ and Lemma \ref{Ann(S|n)} applies to $q_1'$: if $n_i\leq n$ then 
$$\vv(q_1'\cdot\ul{S}-x\,q_2')= 1-|q_1|=1-n_i+ |q_1'|\geq 1+|q_1'|-n$$
i.e.  $q_1'\not\in\Ann(s)$.   We next show that $q_1\in\Min(S|n_i)$.  Since $q_1\in\Ann(s)$ we have  $q_1\in\Ann(S|n_i)$, so $|q_1|\geq \LC_{n_i}$. We have  $n_i\geq n_2>n_1\geq 1$ i.e. $n_i-1\geq 1$. The inductive hypothesis and Lemma \ref{Ann(S|n)}  imply that $q_1'\in\Ann(S|n_i-1)\setminus\Ann(S|n_i)$. So Lemma \ref{onemore2} implies that  $|q_1|\geq \LC_{n_i}\geq n_i-|q_1'|=|q_1|$ and hence $|q_1|=\LC_{n_i}$. Now let $n_i+1\leq n<n_{i+1}$. We know that $q_1\in\Ann(s)$, so $\LC_n\leq |q_1|=\LC_{n_i}$. But $n\geq n_i$ implies that $\LC_{n}\geq \LC_{n_i}$, so $\LC_{n}=\LC_{n_i}$ and  $q_1\in\Min(s)$.

(ii) $\Rightarrow$ (i). If $n\not\in[n_i,n_{i+1})$ either (a) $n\in[n_{i-1},n_i)$ and $q'_1\in\Ann(s)$  or (b) $n\in[n_{i+1},n_{i+2})$ and so $q_1\not\in\Ann(s)$ by the first part, and in particular $q_1\not\in\Min(s)$.
\epr

It follows that for Example \ref{WSexample1}, we have $\LC_1=\LC_2=1$, $\LC_3=\LC_4=2$ and $\LC_5=\LC_6=3$.
 We list some simple consequences of Theorem \ref{simplicityitself}:
  \bp\label{easyconseq} Let $n\geq 1$,  $s=S|n$  be non-trivial. Then
 
 (i) either $s$ is geometric or essential; 
  
 (ii)  $\LC_{n_i}=\sum_{k=1}^i|a_k|$   and we can obtain $\LC_{n_i}$ without computing partial quotients;
 
 (iii) if $n_{i+1}<\infty$ then on the interval $[n_i,n_{i+1})$,  $\LC_{n_i}=|q_1^{(i)}|$ appears $n_{i+1}-n_i$ times.
 \ep
 \bpr We prove (i) only. As $s$ is non-trivial, $n\not\in[1,n_1)$ i.e.  $n\in[n_i,n_{i+1})$ for some $i\geq 1$ and  $n_1=|q_1|=|a_1|=1-\vv$. If $i\geq 2$ then  $\LC_n=\LC_{n_i}\geq \LC_{n_2}>\LC_{n_1}\geq\LC_1$ i.e. $s$ is essential. Now suppose that $i=1$  i.e.  $n\in[n_1,n_2)$. If $\vv=0$ then $|q_1|=1$, $\LC_n=\LC_{n_1}=\LC_1=1$ and $s$ is geometric. If $\vv<0$, $n_1=|q_1|>1$ and $\LC_n=\LC_{n_i}\geq \LC_{n_1}=1-\vv\geq 2$ and $\LC_{n_1}>\LC_1=0$ i.e.  $s$ is essential.
 \epr
We note that in the previous proposition, if  $n\in[n_1,n_2)$ and $b_1=0$ (i.e.  $n_2=\infty$) then $S|n$ is geometric; otherwise  for $k\in[n_i,n_{i+1})$ and $i\geq 2$,  $S|k$ will be essential. The second state diagram of the Introduction illustrates this behaviour.

\bc\label{MasseyT} (Cf. \cite[Theorem 1]{Ma69}) Let $n\geq 1$ and $s=S|n$  be non-trivial. Then

(i) either $\LC_{n+1}=\LC_n$ or  $\LC_{n+1}=n+1-\LC_n>\LC_n$;  

(ii) if $n\in[n_i,n_{i+1})$ and $q_1^{(i)}\not\in\Ann(S|n+1)$ then $\LC_{n+1}=\max\{\LC_n,n+1-\LC_n\}$.
\ec
\bpr  We know that $n\in[n_i,n_{i+1})$ for some $i\geq 1$. (i) From Theorem \ref{simplicityitself}, if  $n+1\in[n_i,n_{i+1})$ then $\LC_{n+1}=\LC_{n_i}=\LC_n$. Otherwise $n+1=n_{i+1}=\LC_{n_i}+\LC_{n_{i+1}}=\LC_n+\LC_{n+1}$ and $\LC_{n+1}=\LC_{n_{i+1}}>\LC_{n_i}=\LC_n$.  (ii) If $\LC_n\geq n+1-\LC_n$ then  $\LC_{n+1}=\LC_n$ by Part (i). Suppose now that $\LC_n<n+1-\LC_n$. Since  $n+1-\LC_n\leq \LC_{n+1}$ by Lemma \ref{onemore2},  $\LC_n< \LC_{n+1}$ and by Part (i) we have $\LC_{n+1}=n+1-\LC_n$. We conclude that  $\LC_{n+1}=\max\{\LC_n,n+1-\LC_n\}$.
\epr

If $S|n$ is essential, the integer  $\max_{1\leq j<n}\{j:\ \LC_j<\LC_n\}$ equals $n_i-1$:
\bc \label{anotherexample}   If  $\LC_n>\LC_1\geq 0$, $n\in[n_i,n_{i+1})$, $s=S|n$, $q=q^{(i)}$, $q'=q^{(i-1)}$ and $n'=n_i-1$ then

(i) $q'$ is a minimal solution for $s'=S|n'$;

(ii) $n'=\max_{1\leq j<n}\{j:\ \LC_j<\LC_n\}$ and $\LC_n+\LC_{n'}=n'+1$;

(iii)  $[q_1\cdot\ul{s}]=xq_2$ and $[q_1'\cdot\ul{s}]=xq_2'$.
\ec
\bpr  We know that $n\in[n_i,n_{i+1})$ for some $i\geq 1$ and $\LC_{n_i}=\LC_n>\LC_1\geq 0$. Hence  $n'\geq 1$ and $s'$ is well-defined.  As $n'\in [n_{i-1},n_i)$,  $q'_1\in\Min(s')$ by Theorem \ref{simplicityitself}. (ii) We have $\LC_{n'}=\LC_{n_i-1}<\LC_{n_i}=\LC_n$ and  so $n'=\max_{1\leq j<n}\{j: \LC_j<\LC_n\}$. Also $\LC_n+\LC_{n'}=|q|+|q'|=n_i=n'+1$.
(iii) We have $[q_1\cdot\ul{s}]=[q_1\cdot\ul{S}]=x\,q_2$ by Lemma \ref{Ann(S|n)} since $|q_1|=\LC_{n_i}\leq n_i\leq n$. Likewise $|q'_1|<|q_1|\leq n$  and   $[q'_1\cdot\ul{s}]=[q'_1\cdot\ul{S}]=x\,q'_2$. 
\epr
We conclude with a  consequence of  Theorem \ref{cfbasics}  and Corollary \ref{anotherexample}.   
\bp \label{cfidentity}   Suppose that $\LC_n>\LC_1\geq 0$, $n\in[n_i,n_{i+1})$, $q=q^{(i)}$ and $q'=q^{(i-1)}$. If 
$f\in\R^2$, $m=f_2 \,q_1-f_1\,q_2$ and $m'=f_2\,q'_1-f_1\,q_2'$ then
 $$(-1)^{i-1}\, f_1= m'\,q_1-m\,\,q_1'.$$ 
\ep
\bpr   As $q'$ is well-defined  from Corollary \ref{anotherexample}, we have  $m'\,q_1-m\,\,q_1'=
(f_2\,q_1'-f_1\,q_2')\,q_1-(f_2\,q_1-f_1\,q_2)\,q_1'=
f_1(q_2\,\,q_1'-q_2\,'q_1)$, which is $(-1)^{i-1}\, f_1$ by Theorem \ref{cfbasics}.
\epr 

  \section{An Inductive Construction of Minimal Solutions}
  \label{myconstr}
  In this section, we work with arbitrary finite sequences over $\D$. Given $n\geq 1$,  we construct   a minimal solution $\mu$  for any $s=s_0,\ldots,s_{1-n}$ i.e. $\mu_1\in\Ann(s)$ and $|\mu_1|=\LC_n=\LC$ say. If $\LC<n$ and $\lc(\mu_1)$ is a unit of $\D$, we can 'generate'  $s_{-\LC},\ldots,s_{1-n}$ using $s_0,\ldots,s_{1-\LC}$ and $\mu_1$. For Examples  \ref{first}(ii), (iii) we will see that the construction returns $\mu_1=x^n$ and $\mu_1=2x-1$ respectively; in each case, we cannot generate $s_{1-n}$ using $s_0,\ldots,s_{2-n}$ and $\mu_1$.
    
We simplify \cite{N95b}  by appealing to Proposition \ref{split} and  recalling two  lemmas from \cite{N95b}. In this way we construct a new minimal solution when the current one fails. The proof of each lemma consists of (i) verifying that we have a new solution and  (ii) applying Lemma \ref{onemore2} to deduce minimality. 
For a pseudo-geometric sequence, it suffices to consider $n+1-2\LC=n-1>0$ only and the proof is elementary. However, for an essential sequence, we require both a current minimal solution and a solution for $s_0,\ldots,s_{1-n'}$ where $1\leq n'<n$. We encode each of these solutions as a 'triple'.

The resulting Algorithm \ref{calPA} is identical to \cite[Algorithm 4.6]{N95b}, except that we compute $\mu_2$ rather that $x\,\mu_2$. 
We can also suppress second components and in this way compute minimal polynomials only, cf. \cite{Ma69}. We  include a normalised version to compute a monic $\mu_1$ when $\D$ is a field. We also give a new analysis of Algorithm \ref{calPA}.

 Section \ref{myconstr}.5 defines the scalar $\nabla_s\in\D^\times$ and proves Identities (\ref{nablaeqn}), (\ref{basic}) of the Introduction, see Propositions \ref{identity}, \ref{fid}. These identities are integral to the rest of the paper.
 \subsection{Pseudo-Geometric Sequences}
The following integer will play an important role for all finite sequences. 
\bd For $n\geq 1$ and $s=s_0,\ldots,s_{1-n}$ we put $e_s=n+1-2\LC_n\in\Z$.
\ed

 \bl \label{s_0=1}(\cite[Theorems 3.8, 4.5]{N95b}) 
Let $k\geq 1$, $r=s_0,\ldots,s_{1-k}$, $s_0\neq 0$ and $\mu$ be a minimal solution for $r$ with $\LC_k=|\mu_1|=1$.  If $s=r,s_{-k}$ and $\Delta_1=\Delta(\mu_1;s)\neq 0$ then $\nu=s_0x^{e_r}\mu-\Delta_1(1,0)$ is a minimal solution for $s$; in fact $|\nu_1|=\max\{1,1+e_r\}$.
\el

We apply this as follows. If $n=1$ and $s_0\neq 0$ then $\mu=(x,s_0)$, a pseudo-geometric sequence and $e_{s_0}=0$. For $n=2$, $r=s_0$ and $s=r,s_{-1}$, if $\Delta_1=s_{-1}\neq 0$ then $\nu=s_0x^0(x,s_0)-s_{-1}(1,0)=(s_0x-s_{-1},s_0^2)$ since $e_r=0$; $s$ is also pseudo-geometric.  Let $n\geq 3$ and $s=r,s_{1-n}$. If  $\LC_{n-1}=\cdots=\LC_1$, $\mu$ is a minimal solution for $r$ and $\Delta_1=\Delta(\mu_1;s)\neq 0$, then $\nu=s_0x^{n-2}\mu-\Delta_1(1,0)$ since $e_r=n-2$.  Now $\LC_n=|\nu_1|=n-1>\LC_1=1$; the new sequence  $s$ is essential. This is an explicit version of Proposition \ref{transition}.

\begin{examples}\label{second}  Let $a,b\in\D^\times$. (i)  Put $r=a$ and  $s=r,b$. We begin with  $\mu=(x,a)$ and $\Delta_1=\Delta(\mu;s)=b\neq 0$. Lemma \ref{s_0=1} shows that $a(x,a)-b(1,0)=(ax-b,a^2)$ is a minimal solution for $s$. If $a$ is a unit of $\D$ and $m=b/a$, we have $s=a,ma$ i.e. Example \ref{first}(i) and since $ax-b\in\Min(s)$, we have $x-m\in\Min(s)$.

(ii) Now let $k=2$, $r=a,b$ and $s=r,c$. We know that $(ax-b,a^2)$ is a minimal solution for $r$ and $$\Delta_1=\Delta(ax-b;s)=(\,(ax-b)\cdot(cx^{-2}+bx^{-1}+a)\,)_{1-2}=ac-b^2.$$ Lemma \ref{s_0=1} implies that if $\Delta_1\neq 0$, $ax(ax-b,a^2)-\Delta_1(1,0)=(a^2x^2-abx-\Delta_1,a^3x)$ is a minimal solution for $s$.
Taking $\D=\Z$, $a=b=1$ and $c=2$, we see that $x^2-x-1\in\Min(1,1,2)$, as expected. Further, if $f_1=x^3$ then $f_2=x^2+x+2$ and $\gcd(f_1,f_2)=1$. Thus the converse of Proposition \ref{minfield}  fails for essential sequences too.

(iii) Let the common multiple $m$ in Example \ref{first}(i) be zero, so that $r=a,0^{k-1}$ where $k\geq 2$ and $\mu=(x,a)$ i.e. $e_r=k-1$. If $s=r,b$ where $b\in\D^\times$ then $\Delta_1=\Delta(x;s)=b\neq 0$, hence $ax^{k-1}(x,a)-b(1,0)=(ax^k-b,a^2x^{k-1})$ is a minimal solution  for $s$ and $\LC_{k+1}=k>1=\LC_k$.
\end{examples}
\subsection{Essential Sequences or, a Tale of Two Triples}
Next we recall a lemma which constructs a minimal solution for an essential sequence when the current one fails. As this is more involved, we encode the data as a 'triple' consisting of a strictly positive integer, a minimal polynomial and an element of $\D^\times$.
 We also require that our two triples are linked by linear complexity. Thus  given a pair of  linked triples for $r$ and $a\in\D$,  we construct a pair of linked triples  for $s=r,a$. 

\bd \label{triple}Let $k\geq 1$, $r=s_0,\ldots,s_{1-k}$ be essential and $s_{-k}\in\D$. {\em A linked pair of triples (for $r$)} consists of  $T_r=(k,\mu_1,\Delta_1),\ T'_r=(k',\mu'_1,\Delta'_1)\in\N^\times\times\R\times\D^\times$ such that

(i)  $s=r,s_{-k}$\,, $\Delta_1=\Delta(\mu_1;s)$,  $\mu_1\in\Min(r)\setminus\Ann(s)$;

(ii) $k'=\max_{1\leq j<k}\{j:\ \LC_j<\LC_k\}<k$ so that  $r'=s_0,\ldots,s_{1-k'}$ is a well-defined,     proper subsequence of $r$;

(iii) $s'=r',s_{-k'}$\,, $\Delta'_1=\Delta(\mu'_1;s')$, $\mu'_1\in\Min(r')\setminus\Ann(s')$  and $\LC_k+\LC_{k'}=k'+1$. 
\ed
\be\label{linkedtriples} Let $a,b\in\D^\times$. 

(i) 
 Let $r=a,0^{k-1}$ with $k\geq 2$  (a pseudo-geometric sequence with common multiple $0$), $s=r,b$ and  $\nu=(ax^k-b,a^2x^{k-1})$ be a minimal solution for $s$ as in Example \ref{second}(iii). Let $T'_s=(k,x,b)$ and suppose that $\Delta_1=\Delta(\nu_1;s,c)\neq 0$ for some $c\in\D$. Then $T_s=(k+1,\nu_1,\Delta_1)$ and $T'_s$ are linked: for $i\leq k$ we have $\LC_i=1$, so $(k+1)'=k$ and $\LC_{k+1}+\LC_{(k+1)'}=k+1=(k+1)'+1$. 

(ii) Let $k\geq 2$ and $r=0^{k-1}, a$ where $k\geq 2$, so $\vv=1-k$. We claim that  $T_r=(k,x^k,b)$ and $T'_r=(k-1,1,a)$  are linked. We have $\Delta(1;0^{k-1})=a\neq 0$, giving the triple $T'_r=(-\vv, 1,a)$ i.e. $k'=-\vv=k-1$ and $\LC_{k'}=0$. We also know that $(x^k,a)$ is a minimal solution for $r$. Let $s=r,b$ where $b\neq 0$. Then 
$$\Delta(x^k;s)=(x^k\cdot\ul{s})_{k-k}=s_{-k}=b\neq 0.$$
giving the triple $T_r=(k,x^k,b)$. Furthermore $T_r,T'_r$ are linked since  $\LC_k+\LC_{k'}=k=k'+1$. 
\eex
\br In \cite{N95b}, we used  $T_r=(1,x,\Delta_1)$,  $T'_r=(0,1,s_0)$, linked by $\LC_0=0$  when $r=s_0\neq 0$.  In this paper, we treat pseudo-geometric sequences separately and the proper subsequence $r'$ always has length  $n'\geq 1$, further simplifying the theory developed in  \cite{N95b}.
\er

We now combine several results from \cite{N95b} to construct a linked triple for $s=r,s_{-k}$ from a linked triple for $r$.
\bl \label{template}(\cite[Proposition 3.11, Theorem 3.13, Proposition 4.4]{N95b}) Let $k\geq 2$, $r=s_0,\ldots,s_{1-k}$  be essential and $s=r,s_{-k}$. Suppose that 
 $T_r=(k,\mu_1,\Delta_1)$, $T'_r=(k',\mu'_1,\Delta_1')$ are linked triples for $r$. If $$ \nu=\left\{\begin{array}{ll}
\Delta_1'\,  \mu-\Delta_1\, x^{-e}\mu'&\mbox{ if }e=e_r\leq 0\\\\
\Delta_1'\,  x^{+e} \mu-\Delta_1\, \mu'&\mbox{ otherwise}\end{array}\right.
$$
then   $|\nu_1|=\max\{\LC_k,\LC_k+e_r\}=\max\{\LC_k,k+1-\LC_k\}$, $\nu_1\in\Min(s)$ and $x\,\nu_2=[\nu_1\cdot t]$ i.e. $\nu$ is a minimal solution for $s$. 
Further, if 

(a) $\Delta=\Delta(\nu_1;s,s_{-k-1})\neq 0$ and $T_s=(k+1,\nu_1,\Delta)$;  

(b) $T'_s=T'_r$ if $e_r\leq 0$ and $T'_s=(k,\mu_1,\Delta_1)$  if $e_r\geq 1$; 

then $T_s,T'_s$ are linked. Finally $e_s=1+e_r$ if $e_r\leq 0$, and $e_s=1-e_r$ otherwise. 
 \el
\bpr  We show only that $T_s,T'_s$ are linked.  (The remaining item on updating $e_s$ is a simple  verification.)
Let $m_k=\max_{1\leq j<k}\{j: \LC_j<\LC_k\}$ and $\mu_1\in\Min(r)$. If $e_r\leq 0$ then $\LC_{k+1}+\LC_{(k+1)'}=\LC_k+\LC_{k'}=k'+1=(k+1)'+1$ and  $(k+1)'=k'=m_k=m_{k+1}$. Otherwise $\LC_{k+1}+\LC_{(k+1)'}=(k+1-\LC_k)+\LC_k=k+1=(k+1)'+1$ and $(k+1)'=k=m_{k+1}$ since $\LC_k<k+1-\LC_k=\LC_{k+1}$.  
\epr

\be\label{linkedexamples} (i) For Example \ref{linkedtriples}(ii),   Theorem \ref{template} yields $\nu_1=ax^k-bx^{k-1}$ from $T_r=(k,x^k,b)$, $T'_r=(k-1,1,a)$ which are linked as we have seen. 
(ii)  In Lemma \ref{s_0=1}, $\LC_k=1$, $\nu_1=s_0x^{k-1}\mu_1-\Delta_1$ and $\LC_{k+1}=k$. If $\Delta=\Delta(\nu_1;s)\neq 0$ we have $T_s=(k+1,\nu_1,\Delta)$, $T'_s=(k,\mu_1,\Delta_1)$ and $\LC_{k+1}+\LC_{(k+1)'}=k+1=(k+1)'+1$ so that $T_s$, $T_{s'}$ are linked.
\eex
 \subsection{The Inductive Theorem and the Corresponding Algorithm}
The elementary case $n=1$, Example \ref{linkedtriples}(ii),  Lemmas \ref{s_0=1} and \ref{template} now yield 
\bt\label{indthm}(\cite[Theorems 3.13, 4.5]{N95b}) For $n\geq 1$ and any  sequence $s=s_0,\ldots,s_{1-n}$  over $\D$, we can construct a minimal solution $\mu$ for $s$. 
\et
\bpr We induct on $n$. For  $n=1$, $\mu=(1,0)$ is  minimal  if $s$ is trivial and otherwise $\mu=(x,s_0)$ is. Let $n=2$, $r=s_0$ and $s=r,s_{-1}$. If $\Delta_1=\Delta(\mu_1;s)=0$ then $\mu$ is as required. Otherwise $\Delta_1\neq 0$ and $s$ is non-trivial, so $s$ is either  pseudo-geometric or essential. In the first case, $r$ is also pseudo-geometric and we can apply Lemma  \ref{s_0=1} to $r$ and $s_{-1}$: we take $\mu=(s_0x-s_{-1},s_0^2)$. For the second case,  $r$ is trivial and $s_{-1}\neq 0$ so we take $\mu=(x^2,s_{-1})$. 
Morover if we put $n'=1$ then $\LC_n+\LC_{n'}=2=n'+1$. Hence if $s$ is non-trivial and $\Delta_1=\Delta(x^2;s,s_{-2})\neq 0$, we have linked triples $T_s=(2,x^2,\Delta_1)$,  $T'_s=(1,1,s_{-1})$.

Now let $n\geq 3$, $s=r,s_{1-n}$ and $\mu$ be our solution for $r$ with linked triples $T_r,T'_r$ if both $\ul{r}\neq 0$ and $\Delta_1=\Delta(\mu_1;s)\neq 0$. Thus $s$ is non-trivial; if $s$ is pseudo-geometric, so is $r$ and we apply Lemma \ref{s_0=1} to $r$ and $s_{1-n}$.  Otherwise $s$ is essential.
 If $\ul{r}=0$ then $s_{1-n}\neq 0$. Put $\mu=(x^n,s_{1-n})$.  Now $n'=n-1$, $\LC_n+\LC_{n'}=n=n'+1$ and $T'_s=(n',1,s_{1-n})$ is a triple. For $\ul{r}\neq 0$, the inductive hypothesis and Lemma \ref{template} apply to $r$, $s_{1-n}$, $\mu$ and  linked triples $T_r,T_r'$ to yield a new $\mu$, and a linked $T_s,T'_s$ if $s$ is non-trivial and $\Delta(\mu_1;s,s_{-n})\neq 0$. 
\epr
\br\label{indthmremarks} The proof of Theorem \ref{indthm}  for minimal polynomials only does not require the fact that $x\mu_2=[\mu_1\cdot \ul{s}]$.
 \er
 
\bc (\cite[Theorem 2]{Ma69}, cf. Corollary \ref{MasseyT}) If $s=s_0,\ldots,s_{1-n}$ is non-trivial and $\mu_1\not\in\Ann(s,s_{-n})$ then   $\LC_{n+1}=\max\{\LC_n,n+1-\LC_n\}$.
\ec
\bpr As $\mu_1\not\in\Ann(s,s_{-n})$, Theorem \ref{indthm} implies that $\nu_1\in\Min(s,s_{-n})$ i.e. $\LC_{n+1}=|\nu_1|$. If $e_s\leq 0$, $\LC_{n+1}=\LC_n\geq n+1-\LC_n$. Otherwise $e_s>0$  and $\LC_{n+1}=n+1-\LC_n>\LC_n$. 
\epr

Next we derive the algorithm which follows from the constructive proof of Theorem \ref{indthm}. The constructions in Example \ref{linkedtriples}(ii) and Lemma \ref{s_0=1} bear some resemblance to Lemma \ref{template}, and pseudo-geometric sequences often become essential. Thus it is reasonable to try to fit these two   cases  into  the format of Lemma \ref{template} and to iterate. 

First we rewrite Lemma \ref{template} algorithmically using $\mu$ for the current solution,  the variable $\Delta_1'$,  updates for $e$  and $\nu$ for the new solution.
Here $2\leq k\leq n-1$ and $s=s_0,\ldots,s_{1-k}$ is essential:\\

\noindent {\sc Lemma} \ref{template} (restated)
\begin{tabbing}
$\Delta_1\la\Delta(g_1;s_0,\ldots,s_{-k})$; \\
{\tt if } $\Delta_1\neq 0$ \={\tt then }\= {\tt if }  $ e\leq 0$ \={\tt then } $\lceil \nu\la(\Delta_1'\,\mu-\Delta_1\, x^{-e}\mu'$; $e\la 1+e;\,\rfloor$ \\
                         \>\>\>{\tt else }  $ \lceil \nu\la \Delta_1' \,x^e \mu-\Delta_1\,\mu'$; 
                         $(\mu',\Delta_1')\la(\mu, \Delta_1)$; $e\la 1-e;\,\rfloor$
\end{tabbing}

 (i) We observe how   Lemma \ref{template} (restated) reduces when $k=0$ if we start with $\mu=(1,0)$, $e=1$, $\mu'=(0,-1)$ and $\Delta_1'=1$: we have $\Delta(\mu_1;s_0)=s_0$ and thus if $s_0\neq 0$ we have  $\nu=\Delta_1'\,x^e\, \mu-\Delta_1\,\mu'=x(1,0)-s_0(0,-1)=(x,s_0)$ and the case $e\leq 0$ does not arise.  We have the correct result when $s_0\neq 0$;  now $e=0$, $\nu\,'=(1,0)$ and $\Delta_1'=s_0$.

(ii) Now let $k=2$ and put $\mu=\nu$, $\mu'=\nu\,'$. We have $\Delta_1=\Delta(\mu_1;s)=s_{-1}$. If  $s_{-1}\neq 0$ and $e=0$ i.e. $s_0\neq 0$ we have  $\nu=\Delta_1'\mu-\Delta_1\, x^{-e}\mu'=s_0(x,s_0)-s_{-1}(1,0)=(s_0x-s_{-1}, s_0^2)$ as desired. But if $s_0=0$  then $s$ is essential and $\mu=(1,0)$, giving $\nu=\Delta_1'\,x^e \mu-\Delta_1\,\mu'=x^2(1,0)-s_{-1}(0,-1)=(x^2,s_{-1})$, $\nu\,'=(1,0)$ and $\Delta_1'=s_{-1}$, provided $e=2$. So  Lemma \ref{template}  (restated) behaves correctly when $k=2$, provided $e=2$ if $s_0=0$. 
  We conclude that  when $\Delta_1\neq 0$,  Lemma \ref{template}   (restated) applies if we initialise as in case (i) and $e\la 1+e$  if $s_0=0$. Moreover $s,s_{-2}$ will be essential, so that Lemma \ref{template}   (restated) can be reapplied.\\
   
Now replace $\nu$ by $\mu$ and $\nu\,'$ by $\mu'$  throughout. This requires a temporary variable $T$ to avoid overwriting $\mu$ when $e>0$. We factor out incrementing $e$, giving last statement   $e\la 1+e$.

If $\Delta_1=0$ then  $\mu_1$ remains unchanged and  $e\la 1+e$ since $k+2-2|\mu_1|=1+e$. We can thus place  $e\la 1+e$ at the bottom of the loop, independently of $\Delta_1$, as in:

 \begin{algorithm}\label{calPA}(\cite[Algorithm 4.6]{N95b}, cf. \cite[Algorithm 1]{Ma69})
\begin{tabbing}
\noindent {\tt Input}: \ \ \=  The $n\geq 1$ values of a sequence $s=s_0,\ldots,s_{1-n}$ over $\D$.\\

\noindent {\tt Output}: \> A minimal solution $\mu$ for $s$.\\\\

$\lceil\, \mu\leftarrow(1,0)$;  $\mu'\leftarrow(0,-1)$; $\Delta_1'\leftarrow1$;\ $e \leftarrow 1$;\\\\

{\tt for} \= $i\leftarrow0$ {\tt to }$1-n$ {\tt do}\\\\

   \>$\lceil \ \Delta_1\leftarrow\Delta(\mu_1;s_0,\ldots,s_i)$;  \\\\
    \>{\tt if}$\ \Delta_1\neq 0$ {\tt then} \={\tt if} \=$e\leq 0$\ \={\tt then}\, $\mu\leftarrow \Delta_1'\, \mu-\Delta_1\,  x^{-e}\, \mu'; $\\\\
  \> \> \>\> {\tt else}\ \=
   $\lceil T\leftarrow\mu$; $\mu \leftarrow \Delta_1'\, x^e\, \mu-\Delta_1\, \mu'$; \\\\
    \>\>\>\>\>$(\mu',\Delta_1')\leftarrow (T,\Delta_1)$;\ $e \leftarrow -e;\rfloor$\\\\
   \> $e  \leftarrow 1+e;\  \rfloor\ $\\\\
{\tt return }$\mu.\rfloor$
\end{tabbing} 
\end{algorithm}
Note that after $s_0=0$ we have $e\geq 2$. We verify the remaining cases:
 
 (iii) $s$ trivial; Algorithm \ref{calPA} gives $\mu=(1,0)$ as it should;
 
 (iv) $s=0^{k-1},s_{1-k}$ where $s_{1-k}\in\F^\times$; here 
 $\mu=(1,0)$, $e=(k-1)+1-2|\mu_1|=k$ and $\Delta(\mu_1;s_0,\ldots,s_{1-k})=s_{1-k}$, Algorithm \ref{calPA} gives $\mu=\Delta_1'\,x^e\,\mu-\Delta_1\,\mu'=x^k(1,0)-s_{1-k}(0,-1)=(x^k,s_{1-k})$, $\mu'=(1,0)$ and $\Delta_1'=s_{1-k}$. This agrees with Example \ref{linkedtriples}(ii) and moreover Lemma \ref{template} can be reapplied.
 
(v) $k\geq 2$ and $s_0,\ldots,s_{1-k}$ is pseudo-geometric; here  $\mu=(s_0\,x- s_{-1}, s_0^2)$, $\mu'=(1,0)$ from (ii) above and $e=k+1-2|\mu_1|=k-1$. If $\Delta_1=\Delta(\mu_1;s_0,\ldots,s_{-k})\neq 0$ then Algorithm \ref{calPA}  gives $\Delta_1'\,x^e \mu-\Delta_1\,\mu'=\Delta_1'\,x^{k-1}\mu-\Delta_1\mu'$ and $\mu=\mu'$, $\Delta_1'=\Delta_1$, which agrees with Lemma \ref{s_0=1}, and Lemma \ref{template} can be reapplied. Finally, if $\Delta_1=0$ then $\mu$ is unchanged. \\

We conclude that  Algorithm \ref{calPA} computes a minimal solution $\mu$ for $s$. 
Note that (i) we may suppress second components and compute $\mu_1$ only as in \cite{Ma69}; (ii) Algorithm \ref{calPA} is identical to \cite[Algorithm 4.6]{N95b} except that  $\mu'\la(0,-x)$ has been replaced by $\mu'\la(0,-1)$,  so that Algorithm 4.6, {\em loc. cit.} computes $x\,\mu_2$ instead of $\mu_2$.  

\br[Initialisation]  \label{arbyd}In \cite[Section 7.3]{Be68}, $(\sigma^{(0)},\omega^{(0)})=(1,1)$ and  $(\tau^{(0)},\gamma^{(0)})=(1,0)$. This corresponds to the fact that $1+s_1x+\cdots +s_nx^n\in\F[x]$  is used in the key equation  \cite[Equation 7.302]{Be68}. Thus  if $s=0^n$ then $1=1/1$ obtains in \cite{Be68}, whereas $0=0/1$ obtains in our approach. 

Our initialisation $\mu'=(0,-1)$ was chosen to yield the inductive bases of Theorem \ref{indthm}. In \cite{Ma69}, we have the initialisation '$B(D)=1$', which corresponds to $\mu'_1=1$. Let Algorithm \ref{calPA}$\,'$ denote Algorithm \ref{calPA} using the initialisation $\mu'_1=1$. The reader may easily check that the first iteration of Algorithm \ref{calPA}$\,'$ (with $\Delta_1=s_\vv\neq 0$) produces   $\mu_1=x^n-s_{\vv}\in\Min(s)$. As Lemmas \ref{s_0=1}, \ref{template} apply to any $\mu_1\in\Min(s)$,  Theorem  \ref{indthm} and hence Algorithm \ref{calPA}$\,'$ also produces a minimal polynomial on subsequent iterations.
\er
\bp[Normalised Algorithm \ref{calPA}]\label{monic} If $\D=\F$ is a field,  $\rho=\Delta_1/\Delta_1'$ and $\mu$ of Algorithm \ref{calPA} is updated via
$$\mu\la\left\{\begin{array}{ll}
 \mu-\rho\,x^{-e}\, \mu'&\mbox{ if } e\leq  0\\
 x^{+e}\mu-\rho\, \mu'& \mbox{ otherwise}
\end{array}
\right.
$$
then  Algorithm \ref{calPA} produces a minimal solution  $\mu$ for $s$ with $\mu_1$ monic. \ep
\bpr It suffices to show that the updating is well-defined and  $\mu_1$ is monic. Firstly,  $\Delta'_1=1$ initially and $\Delta_1'$ is either unchanged or replaced by  $\Delta_1\neq 0$. Thus $\rho$  is well-defined. Secondly, 
$\mu_1$ is monic for the base cases. Suppose that  $s=s_0,\ldots,s_{1-n}$ is essential and $e\leq 0$. Then  $\mathrm{lc}( \mu_1-\rho\, x^{-e}\, \mu'_1)=\mathrm{lc}(\mu_1)$ since $-e+\LC_{n'}=2\LC_n-n-1+\LC_{n'}=\LC_n+n'-n<\LC_n$ as $\LC_n+\LC_{n'}=n'+1$ and $n'<n$. Hence the updated $\mu_1$ will be monic in this case.  And {\em a fortiori} if $n\geq 2$ and $s$ is either (i) geometric or (ii) essential and $e\geq 1$.
\epr
 \begin{table}\label{reallyshortex}
\caption{Algorithm \ref{calPA} for Example \ref{second}(ii)}
\begin{center}
\begin{tabular}{|l|l|l|l|l|l|}\hline
$s$ & $\Delta_1$ & $\Delta'_1$  & $e_s$      			&$\mu$ &$\mu'$\\\hline\hline
$\ $   &$-$   &$1$&$1 $  & $(1,0)$ & $(0,-1)$  \\\hline
$a$   &$a$   &$1$&$1$  & $(x,a)$ &$(1,0)$ \\\hline
$a,b$   & $b$  &$1$&$0$ & $(ax-b,a^2)$ & $(1,0)$  \\\hline
$a,b,c$   & $ac-b^2$  &$a$ &$1$  & $ax(ax-b,a^2)-\Delta_1(1,0)$ & $(ax-b,a^2)$.  \\\hline
\end{tabular}
\end{center}
\end{table}

\be\label{geometric} Let $n\geq 2$ and $s=s_0,\ldots,s_{1-n}$ be a geometric sequence over $\F$ with common ratio $r=s_{-1}/s_0\in\F$.  Proposition \ref{monic} yields iterations $(x,s_0)$ and  $(x-r,s_0)$.
\eex
\begin{example}\label{WSexample2} (Cf. Example \ref{WSexample1}) Let $s_i=\ul{S}_{\,i}$ where $\ul{S}$ is as in Example \ref{WSexample1}. As in \cite{WS}, $s=\alpha^{5},\alpha^{9},\alpha^4, 0,0,\alpha^{2}$. Normalising Algorithm \ref{calPA}   gives the following table: 
\begin{center}
\begin{tabular}{|l|c|c|l|l|}\hline
$s$ &$\Delta_1$  &$\Delta'_1$& $e_s$&$\mu$\\\hline\hline
$-$ &$-$  &$1$  &$1$&$(1,0)$\\\hline\hline
$\alpha^{5}$ &$\alpha^5$ &$1$ &$1$&$(x,\alpha^{5})$\\\hline
$\alpha^{5},\alpha^{9}$ &$\alpha^9$  &$\alpha^{5}$&$0$& $(x+\alpha^4,\alpha^{5})$\\\hline\hline
$\alpha^{5},\alpha^{9},\alpha^4$  &$\alpha^{11}$&$\alpha^{5}$&$1$ &$(x^2+\alpha^{4}x+\alpha^6,\alpha^{5}x)$ \\\hline
$\alpha^{5},\alpha^{9},\alpha^4,0$  &$\alpha^{2}$ &$\alpha^{11}$&$0$&$(x^2+\alpha^{12}x+\alpha^{7},\alpha^{5}x+\alpha^{11})$  \\\hline\hline
$\alpha^{5},\alpha^{9},\alpha^4, 0,0$  &$\alpha^{11}$ &$\alpha^{11}$&$1$& $(x^3+\alpha^{12}x^2+\alpha^{9}x+\alpha^4,\alpha^{5}x^2+\alpha^{11}x+\alpha^5)$ \\\hline
$\alpha^{5},\alpha^{9},\alpha^4, 0, 0,\alpha^2$ &$0$ &$\alpha^{11}$ &$0$ &$(x^3+\alpha^{6}x^2+\alpha^3x+\alpha^{13},\alpha^{5}x^2+\alpha^2x+\alpha^{10})$.\\\hline\hline
\end{tabular}
\end{center}
For $r=s_0\neq0$, $\mu'(r)=\mu'(r,s_{-1})=(1,0)$; for $-5\leq j\leq -2$ and $r=s_0,\ldots,s_{j}$\,,  $\mu'(r)=\mu'(r,s_j)=\mu(s_0,\ldots,s_{2+j})$.
Here  $\LC_1=\LC_2=1$, $\LC_3=\LC_4=2$ and $\LC_5=\LC_6=3$,  which agrees with Theorem \ref{simplicityitself}. 
Note that when $n=2,4,6$ in this table and $i=1,2,3$ in Example \ref{WSexample1}, $\mu= q^{(i)}/\lc(q_1^{(i)})$.
\eex

\subsection{A Worst-Case Analysis}
Next we give a worst-case analysis of Algorithm \ref{calPA}.
For $n\geq 1$ and $s=s_0,\ldots,s_{1-n}$ define $\sigma_n=\sum_{i=0}^{1-n}\LC(s_0,\ldots,s_i)$. The  following identity and inequality  seem to be new.

\bp\label{stable} If $s=s_0,\ldots,s_{1-n}$ then  $\sigma_n=\LC_n(n+1-\LC_n)\leq (n+1)^2/4$, with equality if and only if $n=2\LC_n-1$.
\ep
\bpr   The equality is trivially true if $s=0^n$.  For the sequence $s=0^{n-1},s_\vv$ with $-\vv\geq 0$, we have $\sigma_n=n=n(n+1-n)$ as required. 
Suppose inductively that $n\geq 2$, $s$ is non-trivial, the equality is true for $s=s_0,\ldots,s_{1-n}$ and $t=s,s_{-n}$.  If $\Delta(\mu_1;t)=0$ then $\LC_{n+1}=\LC_n$ and by the inductive hypothesis 
$$\sigma_{n+1}=\sigma_n+\LC_n=\LC_n(n+1-\LC_n)+\LC_n=\LC_n(n+2-\LC_n)=\LC_{n+1}(n+2-\LC_{n+1}).$$ If $\Delta(\mu_1;t)\neq 0$ we apply Lemma \ref{s_0=1} or \ref{template}. If $n+1-\LC_n\leq \LC_n$ then $\LC_{n+1}=\LC_n$ and we have just seen that the result is true in this case. If $n+1-\LC_n>\LC_n$ then $\LC_{n+1}=n+1-\LC_n$ and by the inductive hypothesis, 
$$\sigma_{n+1}=\sigma_n+(n+1-\LC_n)=\LC_n(n+1-\LC_n)+(n+1-\LC_n)=(\LC_n+1)(n+1-\LC_n)$$
Secondly, the right-hand side is $(n+1-\LC_n)(n+2-(n+1-\LC_n))=(n+1-\LC_n)(\LC_n+1)$ which we have just seen is $\sigma_{n+1}$. This completes the inductive proof of  equality.

For the inequality, we show that
$4\LC_n(n+1-\LC_n)\leq (n+1)^2$.  For  integers $a,b$ we have $4ab\leq (a+b)^2$, with equality if and only if $a=b$. Put $a=\LC_n$ and $b=n+1-\LC_n$. Then $a+b=n+1$, so that $4ab\leq (n+1)^2$, with equality if and only if $\LC_n=n+1-\LC_n$.
\epr
\bc (Cf. \cite{Gus76}) Let $s=s_0,\ldots,s_{1-n}$ be a sequence over $\D$. Ignoring  terms linear in $n$, the number of multiplications in Algorithm \ref{calPA} to compute $\mu_1\in\Min(s)$ or a minimal solution $\mu$ for $s$ is at most $c\,n^2/4$ where $c$ is given by

\begin{center}
\begin{tabular}{|l|l|l|}\hline
$\D$& outputs & $c$\\\hline\hline
domain & $\mu_1$& $ 3$\\\hline\hline
domain & solution $\mu$&$ 5$\\\hline\hline
field & monic $\mu_1$ &$2 $\\\hline\hline
field & solution $\mu$, with monic $\mu_1$ &$ 3$.\\\hline\hline\end{tabular} 
\end{center}
\ec 
\bpr For $1\leq k\leq n-1$, let $r=s_0,\ldots,s_{1-k}$ and $\mu_1\in\Min(r)$.  Then $\Delta(\mu_1;r,s_{-k})$ requires at most $\LC_k+1$ multiplications and  $\nu_1\in\Min(r,s_{-k})$ requires at most  $\LC_{k}+1$ if $r$ is pseudo-geometric and  $\LC_{k}+\LC_{k'}+2$ otherwise. If $r$ is essential then by construction
$\LC_{k'}<\LC_{k'+1}=\cdots=\LC_{k}$
so that $\nu_1$ requires at most $3\LC_{k}+2$ multiplications. Thus  computing a minimal polynomial for $s=s_0,\ldots,s_{1-n}$ requires at most 
 $\sum_{k=1}^{n-1} (3\LC_{k}+2)\leq 3n^2/4+2n$ multiplications by Proposition \ref{stable}.  If $(\mu_1,x\mu_2)$ is a solution for $r$ then $|\mu_2|\leq \LC_{k}-1$ and $\mu_2'=0$ or $|\mu'_2|\leq \LC_{k'}-1$, so that we need at most $\LC_{k}+\LC_{k'}$ additional multiplications to obtain $\nu_2$. Ignoring linear terms, this gives  at most $5n^2/4$ multiplications to obtain a solution for $s$. The remaining cases are similar.
\epr
\subsection{An Identity for $\mu$ and $\mu'$}
We prove an identity satisfied by  $\mu,\mu'$. This is our analogue of Identity (\ref{cfid}) satisfied by partial quotients; see Theorem \ref{cfbasics}. First a non-zero scalar:

\begin{definition}\label{nabla} We define $\nabla_s\in\D^\times$ using Algorithm \ref{calPA} as follows: $\nabla_s=1$ on initialisation. Let $\mu_1\in\Min(s)$ and  $t=s,a$. If   $\Delta_1=\Delta(\mu_1;t)=0$ put  $\nabla_t=\nabla_s$; otherwise
$$\nabla_t= \left\{\begin{array}{ll}
\Delta_1'\,\nabla_s
   & \mbox{if }   e_s\leq 0\\
\Delta_1\, \nabla_s& \mbox{otherwise}.
\end{array}
\right.
$$
\end{definition}
If $\D$ is the field of two elements then $\nabla_s=1$ for any $s$.
Suppose that $s=a,b,c$ with $a,b\in\D^\times$ and $c\in\D$ as in Example \ref{second}. After the first iteration, $\mu=(x,a)$, $\mu'=(1,0)$ and $\nabla_a=a$. Next $\mu=(ax-b,a^2)$, $\mu'=(1,0)$ and $\nabla_{a,b}=\Delta_1'\nabla_a=a^2$ since $e_{a,b}=0$. If $\Delta_1=ac-b^2\neq 0$  then  $e_{a,b,c}=1$ and $\nabla_{a,b,c}=\Delta_1\nabla_{a,b}=(ac-b^2)a^2$.

\bp \label{identity}(Cf. \cite[Theorem 7.42]{Be68}) If $\mu$, $\mu'$ are as in Algorithm \ref{calPA} then
$$\mu_2\,\mu_1'-\mu_1\,\mu_2'=\nabla_s.$$ 
\ep

{\sc Proof.}  If $s$ is trivial, $\mu=(1,0)$, $\mu'=(0,-1)$ and $\mu_2\mu_1'-\mu_1\mu_2'=0\cdot 0-1\cdot(-1)=1$.  Suppose inductively that $\mu_2\mu_1'-\mu_1\,\mu_2'=\nabla_s$  and $t=s,a$. If $\Delta_1=0$, there is nothing to prove.  Otherwise let $e_s\leq 0$.  
 By construction $\nu'=\mu'$ and 
\begin{eqnarray*}
\nu_2\,\nu_1'-\nu_1\,\nu_2'&=&(\Delta_1' \mu_2-\Delta_1 x^{-e}\mu_2')\ \nu_1'-
(\Delta_1'\mu_1-\Delta_1 x^{-e}\mu_1')\ \nu_2'\\
&=&(\Delta_1' \mu_2-\Delta_1 x^{-e}\mu_2') \ \mu_1'-
(\Delta_1'\mu_1-\Delta_1 x^{-e}\mu_1')\ \mu_2'\\
&=&\Delta_1'\ ( \mu_2\mu_1'-\mu_1\mu_2')=\Delta_1'\nabla_s=\nabla_t\end{eqnarray*}
whereas if $e_s\geq 1$ we have $\nu'=\mu$ and by construction\begin{eqnarray*}
\nu_2\,\nu_1'-\nu_1\,\nu_2'&=&(\Delta_1' x^{+e} \mu_2-\Delta_1 \mu_2')\ \nu_1'-
(\Delta_1' x^{+e}\mu_1-\Delta_1\mu_1')\ \nu'_2\\
&=&(\Delta_1'  x^{+e}\mu_2-\Delta_1\mu_2') \ \mu_1-
(\Delta_1' x^{+e}\mu_1-\Delta_1 \mu_1')\ \mu_2\\
&=&\Delta_1\ ( \mu_2\mu_1'-\mu_1\mu_2')=\Delta_1\nabla_s=\nabla_t.\ \ \ \square\end{eqnarray*}

If   $\Delta_1=ac-b^2\neq 0$ in  Example \ref{second}(ii), we have seen that $\nabla_{a,b,c}=a^2\Delta_1$ and
$$\mu_2\mu_1'-\mu_1\mu_2'=a^3x\,(ax-b)-(a^2x^2-abx-\Delta_1)a^2=a^2\Delta_1.$$ 
We have the following immediate consequence of Proposition \ref{identity}.
\bc \label{gcdcor}If $s$ is a finite sequence over $\F$ then  $\gcd(\mu_1,\mu_2)=\gcd(\mu_1,\mu_1')=\gcd(\mu_2,\mu_2')=1$.
\ec

The next useful consequence of Proposition \ref{identity} is worth stating separately. The proof is similar to that of Proposition \ref{cfidentity} and is omitted.

\bp \label{fid}
 Let $f\in\R^2$. If $m=f_2\,\mu_1-f_1\mu_2$ and $m'=f_2\mu'_1-f_1\mu_2'$ then $$\nabla_s\, f_1= m'\mu_1-m\,\mu_1'.$$ 
\ep
\begin{example}For $s=0^{n-1},a$ as in Example \ref{first}(ii), $\mu=(x^n,a)$, $\mu'=(1,0)$ and $\nabla_s=a$. For $f\in\R^\times$, $m=f_2\,\mu_1-f_1\mu_2=f_2\,x^n-f_1a$,  $m'=f_2\mu'_1-f_1\mu_2'= f_2$ and
$$m'\mu_1-m\mu'_1=f_2\,x^n-(f_2\,x^n-f_1\,a)=a\,f_1=\nabla_s\,f_1.$$ 
\end{example}

\section{Decomposition}
\label{decomp1}
We now turn to  the set of annihilating polynomials of a finite sequence $s$ over $\D$ (which may be $S|n$ for some infinite sequence $S$ over a field). 

We will characterise the annihilating polynomials which  uses a pairing $\R^2\times\R^2\ra\R$. This pairing was suggested by Identities (\ref{nablaeqn}) and (\ref{basic}) of the Introduction. Even though our conclusions for pseudo-geometric sequences turn out to be a special case of those for essential sequences, we have treated each case separately as their proofs differ, and little would be gained by combining their proofs in one place. Moreover the simpler pseudo-geometric case acts as a precursor to the remaining case.
For essential sequences, the integer $n'$ and the identity $\LC_n+\LC_{n'}=n'+1$ are   vital. 
In each case, we characterise the elements of $\Ann(s)$ using the pairing and show that if we restrict to annihilators of degree at most $n$, our decomposition is unique and we can describe the set of solutions. 

These proofs are valid once we know either a minimal polynomial or a 'minimal system' (see Definition \ref{minsys}) for the original finite sequence i.e. they do not depend on the provenance of the minimal polynomial.
\subsection{A Pairing}
Propositions \ref{cfidentity} and \ref{identity} suggest the following definition:

\bd\label{pairing} For a sequence $s$ we define a pairing $\langle \ ,\ \rangle=\langle \ , \ \rangle_s:\R^2\times \R^2 \ra \R$ by $$\langle f,g\rangle=f_2\,g_1-f_1\,g_2$$
where $x\,f_2=[f_1\cdot\ul{s}]$ and similarly for $g_2$.
\ed

For $s, g,t$ and $h$ as in Lemma \ref{onemore2}, the proof of Lemma \ref{onemore2} shows that $\langle g,h\rangle_t\neq 0$ and $|\langle g,h\rangle_t|=|g|+|h|-n-1\geq 0$. 

From Corollary \ref{anotherexample} and Proposition \ref{cfidentity} we have $(-1)^{i-1} f_1= \langle f,q'\rangle q_1-\langle f,q\rangle\,q_1'.$ We can restate Proposition \ref{identity} as $\langle \mu,\mu'\rangle=\nabla_s$ and Proposition \ref{fid} as  $$\nabla_s f_1= \langle f,\mu'\rangle\mu_1-\langle f,\mu\rangle\,\mu_1'.$$ 
\subsection{Geometric Sequences. II}
\label{PSGII}
Throughout this subsection, $n\geq 1$ and $s=s_0,\ldots,s_{1-n}$ is a pseudo-geometric sequence over $\D$.
We assume that  $\lambda\in\R^2$ is a minimal solution for $s$,  $\LC_n=\cdots=\LC_1=|\lambda_1|=1$ and $\lambda'=(1,0)$, so that $\lambda_2=\langle \lambda,\lambda'\rangle=\nabla\in\D^\times$. 
For example,  if $\lambda$ is obtained via Proposition \ref{cf0} then $\lambda=((x-r)/S_0,1)$ and $\nabla=1$. If $\lambda$ is obtained from Theorem \ref{indthm} then $\lambda_1=x$ and $\nabla=s_0$ if $s_{-1}=0$; otherwise $\lambda_1=s_0x-s_{-1}$ and $\nabla=s_0^2$. In both cases we have
$\nabla f_1=\langle f,\lambda'\rangle \lambda_1-\langle f,\lambda\rangle \lambda_1'.$
\subsubsection{Annihilating Polynomials}
\bl \label{PSGSann} Let $f_1\in\Ann(s)^\times$ and $x\,f_2=[f_1\cdot\ul{s}]$.  If $\varphi\in \R$, $g_1=f_1-\varphi$,  $x\,g_2=[g_1\cdot\ul{s}]$  and $|\varphi|\leq |f_1|-n$ then (i) $|g_1|=|f_1|$, $g_1\in\Ann(s)^\times$ and $g_2=f_2$;  (ii) $\langle g,f\rangle=\varphi f_2 $. 
\el
\bpr (i) Firstly $|\varphi|\leq |f_1|-n\leq |f_1|-1$, so $|g_1|=|f_1|$.   
Since $f_1\in\Ann(s)$ we can write $f_1\cdot \ul{s}=F+x\,f_2$ where $\vv(F)\leq |f_1|-n=|g_1|-n$. Then 
 $$g_1\cdot\ul{s}=(f_1-\varphi)\cdot\ul{s}=F+x\,f_2
-\varphi\cdot\ul{s}$$ and $\vv(\varphi\cdot \ul{s})=|\varphi|+\vv=|\varphi|\leq  |f_1|-n=|g_1|-n$ since $\LC_1=1$ implies that $\vv=0$. Therefore $\vv(g_1\cdot\ul{s}-xf_2)\leq |g_1|-n$, $g_1\in\Ann(s)^\times$ and $g_2=f_2$.  (ii) From Part (i) and the definition $\langle g,f\rangle=g_2f_1-g_1f_2=f_2f_1-(f_1-\varphi) f_2=\varphi f_2$.
\epr
 \bp \label{PSGannconverse} Let  $f\in\R^2$  and $x\,f_2=[f_1\cdot\ul{s}]$. If $m=\langle f,\lambda\rangle$ and  $m'=\langle f,\lambda'\rangle$ then $m=f_2\lambda_1-\nabla\,f_1$,  $m'=f_2$ and $|f_2|=|f_1|-1$. Further   
 
(i) $f_1\in\Ann(s)^\times$ if and only if $|m'|= |f_1|-1$ and $|m|\leq |f_1|-n$;
 
 (ii) if $f_1\in\Ann(s)^\times$ and  $|f_1|\leq n$ then $m\in\D$. 
 \ep
\bpr Since $\lambda_2=\nabla$,  $\lambda_1\cdot\ul{s}=M+x\,\nabla$ where $\vv(M)\leq 1-n\leq 0$. (i) For any $f_1\in\R^\times$,  $m=\langle f,\lambda\rangle=f_2\lambda_1-f_1\nabla\,$ and $m'=\langle f,\lambda'\rangle=f_2$. We have $|m'|=|f_2|=\vv+|f_1|-1=|f_1|-1$. 
Let $f_1\in\Ann(s)^\times$ so that   $f_1\cdot\ul{s}=F+x\,f_2$ where $\vv(F)\leq |f_1|-n$.  Then 
$$f_1\cdot(M+x\,\nabla)=f_1\cdot(\lambda_1\cdot\ul{s})=\lambda_1\cdot(f_1\cdot \ul{s})=\lambda_1\cdot ( F+x\,f_2)$$
$x\,m=x\,(f_2\lambda_1-\nabla\,f_1)=f_1\cdot M-\lambda_1\cdot F$ and $|m|+1\leq\max\{\vv(f_1\cdot M),\vv(\lambda_1\cdot F)\}\leq |f_1|+1-n$ as claimed.   
Conversely, suppose that $|m'|=|f_1|-1$ and $|m|\leq |f_1|-n$. We claim that $m'\lambda_1\in\Ann(s)$: 
$$(m'\lambda_1)\cdot\ul{s}=m'\cdot (M+\nabla\,x)=m'\cdot M+x\,\nabla\,m'$$
and $\vv(m'\cdot M)\leq|m'|+1-n=|m\lambda_1|-n$.  We have $m=f_2\lambda_1-\nabla f_1=m'\lambda_1-\nabla f_1$, so $\nabla f_1=m'\lambda_1-m$. 
Consider $g_1=\nabla\, f_1$: $|g_1|=|f_1|$ and $|m|\leq |f_1|-n=|m'|+1-n=|m'\lambda_1|-n$ since $|m|\leq |f_1|-n=|g_1|-n$. So $g_1\in\Ann(s)$ by Lemma \ref{PSGSann}  and   hence  $f_1\in\Ann(s)^\times$. (ii) This is immediate. 
 \epr
  \bc \label{1mp}
  
  (i)  $\nabla\,\Ann(s)^\times\subseteq\{\varphi' \lambda\,_1-\varphi:\  \varphi'\neq 0, \ |\varphi |\leq |\varphi'|+1-n\}\subseteq\Ann(s)$;
  
  (ii) $\nabla\,\Min(s)\subseteq\{\varphi' \lambda\,_1-\varphi:\  \varphi'\in\D^\times, \ |\varphi |\leq 1-n\}\subseteq\Min(s)$;
  
  (iii) if $n\geq 2$ then $\nabla\,\Min(s)\subseteq\{\varphi' \lambda\,_1:\  \varphi'\in\D^\times\}\subseteq\Min(s)$.
\ec
\bpr (i) If $f_1\in \Ann(s)^\times$ then $\nabla\,f_1=m'\lambda\,_1-m$ where $|m'|=|f_2|=|f_1|-1\geq 0$ and $|m|\leq |f_1|-n=|m'|+1-n$ by Proposition \ref{PSGannconverse}. If   $\varphi'\neq 0$ and $|\varphi |\leq |\varphi'|+1-n$ then $\varphi' \lambda\,_1-\varphi\in\Ann(s)^\times$ by Lemma \ref{PSGSann}. (ii), (iii) These are immediate. 
\epr

\subsubsection{Solutions}
We apply the results of the previous subsection to finding solutions for a pseudo-geometric sequence; this is a precursor to the discussion of solutions for essential sequences in Subsection \ref{esoln}.
\bl\label{PSGphiphi'}Let  $\varphi,\varphi'\in\R$. If $g_1=\varphi'\lambda_1-\varphi$, $0\leq |\varphi'|\leq n-1$, $\varphi\in\D$ and $x\,g_2=[g_1\cdot\ul{s}]$ then $g_2=\varphi'\lambda_2= \nabla\,\varphi'$, $\nabla\, \varphi=\langle g,\lambda\rangle$ and  $\nabla\,\varphi'=\langle g,\lambda'\rangle$.
 \el
 \bpr We have $\lambda_1\cdot\ul{s}=M+x\,\lambda_2$ where  $\vv(M)\leq 1-n$ and
$$g_1\cdot\ul{s}=\varphi'\cdot(\lambda_1\cdot\ul{s})-\varphi\cdot\ul{s}=\varphi'\cdot (M+x\,\lambda_2)-\varphi\cdot\ul{s}=\varphi'\cdot M-\varphi\cdot\ul{s}+x\,\varphi'\lambda_2.$$ 
 Further, $\vv(\varphi'\cdot M)\leq |\varphi'|+1-n\leq 0$ by  hypothesis and $\vv(\varphi\cdot \ul{s})=\vv=0$ since $\varphi\in\D$.   Thus  $\vv(\varphi'\cdot M -\varphi\cdot \ul{s})\leq 0$ and $g_2=\varphi'\lambda_2=\nabla\,\varphi'$. We have $$\langle g,\lambda\rangle=g_2\,\lambda_1-g_1\lambda_2=\nabla\,\varphi'\lambda_1-(\varphi'\,\lambda_1-\varphi)\nabla\,=\nabla\,\varphi$$ and   
$\langle g,\lambda'\rangle=g_2\lambda'_1-g_1\lambda'_2=g_2=\nabla\, \varphi'$.
\epr
\bc \label{1sol} Let $f_1\in\Ann(s)$, $|f_1|\leq n$ and $m'=\langle f,\lambda'\rangle$, $m=\langle f,\lambda\rangle$.  Then

(i) (uniqueness) if   $\nabla\, f_1=\varphi'\lambda_1-\varphi$ where $0\leq |\varphi'|\leq n-1$ and $\varphi\in\D$ then  $\varphi=m$ and $\varphi'=m'$;

(ii) $\nabla\, f_2=m'\lambda_2-m \lambda'_2$. 
\ec
\bpr (i) Applying Lemma \ref{PSGphiphi'} to $g_1=\nabla\,  f_1=\varphi'\lambda_1-\varphi$ gives $\nabla\,\varphi=\langle g,\lambda\rangle=\nabla \langle f,\lambda\rangle=\nabla\,m$. Hence $\varphi=m=f_2$ and similarly  $\varphi'=m'=f_2\lambda_1-\nabla\,f_1$.  (ii) We have $\nabla\,f_2=m'\lambda_2-m \lambda'_2$ since $m'=f_2$, $\lambda_2=\nabla\,$ and $\lambda'_2=0$.  
\epr

Thus if $s$ is a geometric sequence over $\F$, $\mu$, $\nabla$ are obtained from Algorithm \ref{calPA} and $\mu'=(1,0)$  
 then the minimal solutions for $s$ are $$\{(\varphi' \mu-\varphi,\varphi'):\ \varphi'\in\F^\times, |\varphi|\leq 1-n\}.$$

Also if $S_0\neq 0$ and $n\in[1,n_2)$ we may take $\lambda=(a_1,1)$, $\lambda'=(1,0)$ and $\nabla=1$.  For example, from Corollary \ref{1mp}
\bc \label{1cfmp} Let $S$ be an infinite geometric sequence over $\F$, $r=S_{-1}/S_0$, $n\in[1,n_2)$ and $s=S|n$. The minimal solutions for $s$ are $$\{(\varphi' (x-r)-\varphi,\varphi'):\ \varphi'\in\F^\times, |\varphi|\leq 1-n\}.$$
\ec

\subsection{Essential Sequences. II} 
When $s$ is essential, more information is available for decomposition.  Informally, we have a pair of linked triples, their second components and $\nabla\in\D^\times$, all related by the pairing of Definition \ref{pairing}.
\bd\label{minsys} Let $n\geq 2$ and $s=s_0,\ldots,s_{1-n}$ be an essential sequence over $\D$. A {\em minimal system} for $s$ is a 5-tuple $(\lambda,n',\lambda\,',\langle\, ,\,\rangle_s,\nabla)$ consisting of 

(i) a minimal solution $\lambda\in\R^2$ for $s$ and $\lambda_1\not\in\Ann(s,s_{-n})$;

(ii)  $n'=\max_{1\leq j<n}\{j:\ \LC_j<\LC_n\}$ and  $s'=s_0,\ldots, s_{1-n'}$\,;

(iii)   a minimal solution $\lambda\,'\in\R^2$ for $s'$, $\lambda_1'\not\in\Ann(s',s_{-n'})$ and   $\LC_n+\LC_{n'}=n'+1$;

(iv)  the pairing $\langle\, ,\, \rangle_s:\R^2\times\R^2\ra\R$ of Definition \ref{pairing};

(v)  $\nabla=\langle \lambda,\lambda\,'\rangle_s\in\D^\times$.
\ed 

From Theorem \ref{cfbasics} and Corollary  \ref{anotherexample}, 
$(q,n_i-1,q',\langle\, ,\, \rangle_{S|n}\ ,(-1)^{i-1})$
is a minimal system for $S|n$ if $n\in[n_i,n_{i+1})$ and $S|n$ is essential.
We have seen that if  $(n,\mu_1,\Delta)$ and $(n',\mu'_1,\Delta')$ are linked triples for $s$ then $(\mu,n',\mu',\langle\, ,\, \rangle_s\,,\nabla_s)$  is a minimal system for $s$ by  Theorem \ref{indthm}, Definition \ref{nabla} and Proposition \ref{identity}.\\
 
{\bf N.B.} Throughout this subsection,  $n\geq 2$, $s$ is a sequence over $\D$ and $(\lambda,n',\lambda\,',\langle\ ,\ \rangle_s,\nabla)$ is a minimal system for $s=s_0,\ldots,s_{1-n}$. We put $\langle\ ,\ \rangle=\langle\ ,\ \rangle_s$, $\LC=\LC_n$ and $\LC'=\LC_{n'}$. \\

As we have already seen in Propositions \ref{cfidentity} and \ref{identity}(ii),  for any $f\in\R^2$ we have $$\nabla\, f_1=\langle f,\lambda\,'\rangle\lambda_1-\langle f,\lambda\rangle \lambda'_1.$$  
\subsubsection{Annihilating Polynomials}
\bl \label{generalann}  If $f_1\in\Ann(s)^\times$,  $\varphi\in \R$, $g_1=f_1-\varphi\lambda\,'_1$  and $|\varphi|\leq |f_1|+\LC-n-1$ then 

(i) $|g_1|=|f_1|$, $g_1\in\Ann(s)^\times$ and $g_2=f_2-\varphi \lambda\,'_2$;

(ii) $\langle g,f\rangle=\varphi\langle f,\lambda\,'\rangle$. 
\el
\bpr (i) Firstly $|g_1|=|f_1|$ since  $n'<n$ implies that $$|\varphi\lambda\,'_1|=|\varphi|+\LC'\leq |f_1|+\LC-n-1+\LC'=|f_1|-n+n'\leq |f_1|-1.$$
Since $f_1\in\Ann(s)$ we can write $f_1\cdot \ul{s}=F+x\,f_2$ where $\vv(F)\leq |f_1|-n=|g_1|-n$ and $\lambda\,'_1\cdot\ul{s}'=M'+x\,\lambda\,'_2$ where  $\vv(M')\leq \LC'-n'=1-\LC$.
Put $N'=\lambda\,'_1\cdot(\ul{s}-\ul{s}')$. Then $\vv(N')\leq \LC'-n'=1-\LC$ and so $\lambda\,_1'\cdot\ul{s}=N'+M'+x\,\lambda\,'_2$ where $\vv(N'+M')\leq 1-\LC$. Thus  
 $$g_1\cdot\ul{s}=(f_1-\varphi\lambda\,'_1)\cdot\ul{s}=F+x\,f_2
-\varphi\cdot (N'+M'+x\,\lambda\,'_2)=F-\varphi\cdot (N'+M')+x\,(f_2-\varphi\lambda\,'_2).$$
Now  $\vv(\varphi\cdot (N'+M'))\leq |\varphi|+1-\LC\leq |f_1|+\LC-n-1+1-\LC=|f_1|-n=|g_1|-n$, $g_1\in\Ann(s)^\times$ and $g_2=f_2-\varphi\lambda\,'_2$.  
(ii) From Part (i) 
$\langle g,f\rangle=(f_2-\varphi\lambda\,'_2)f_1-(f_1-\varphi\lambda\,_1')f_2=\varphi\langle f,\lambda\,'\rangle.$
\epr
 \bt \label{annconverse} Let $f\in\R^2$. If  $m=\langle f,\lambda\,\rangle$ and  $m'=\langle f,\lambda\,'\rangle$ then  
 
 (i) $f_1\in\Ann(s)^\times$ if and only if $|m'|=|f_1|-\LC$ and $|m|\leq |f_1|+\LC-n-1$;
 
 (ii) if $f_1\in\Ann(s)^\times$ then $|m|+\LC'\leq |f_1|-1$. If in addition $|f_1|\leq n$ then $|m'|\leq n-\LC$ and 
 $|m|\leq \LC-1$. 
 
\et
\bpr First  write $\lambda\,_1\cdot\ul{s}=M+x\,\lambda\,_2$ where $\vv(M)\leq \LC-n\leq 0$.
 (i) Let $f_1\in\Ann(s)^\times$, so that   $f_1\cdot\ul{s}=F+x\,f_2$ where $\vv(F)\leq |f_1|-n$.  Then 
$$f_1\cdot (M+x\,\lambda\,_2)=f_1\cdot (\lambda\,_1\cdot\ul{s})=\lambda\,_1\cdot (f_1\cdot \ul{s})=\lambda\,_1\cdot ( F+x\,f_2)$$
$x\,m=x\,(f_2\lambda\,_1-f_1\lambda\,_2)=f_1\cdot M-\lambda\,_1\cdot F$ and $|m|+1\leq\max\{|f_1|+\vv(M),\LC+\vv(F)\}\leq |f_1|+\LC-n$.  Hence
$$|m|+\LC'\leq |f_1|+\LC-n-1+\LC'= |f_1|+n'-n\leq |f_1|-1$$ as $n'<n$. We have $\nabla\,  f_1=m'\lambda\,_1-m\lambda\,'_1$, so $|m'|=|f_1|-\LC$. Conversely, if $|m'|=|f_1|-\LC$ then $m'\lambda\,_1\in\Ann(s)$, for
$$(m'\lambda\,_1)\cdot\ul{s}=m'\cdot (M+x\,\lambda\,_2)=m'\cdot M+x\,m'\lambda\,_2$$
and $\vv(m'\cdot M)\leq| m'|+\LC-n=|m'\lambda\,_1|-n$. 
We claim that $\nabla\, f_1=m'\lambda\,_1-m\lambda\,'_1\in\Ann(s)$.  From Lemma \ref{generalann}, it suffices to check that $|m|\leq |m'\lambda_1|+\LC-n-1$. But $|m'|=|f_1|-\LC$, so $|m|\leq |f_1|+\LC-n-1$ suffices, and this is true by hypothesis.  We conclude that $\nabla f_1\in\Ann(s)$ and hence so is $f_1$. 
(ii) The first sentence was proved in Part (i); we also have $|m'|=|f_1|-\LC\leq n-\LC$ and $|m|\leq |f_1|+\LC-n-1\leq\LC-1$.
 \epr
 
 Recall that for any sequence $s=s_0,\ldots,s_{1-n}$ over $\D$, $e_s=n+1-2\LC_n\in\Z$.
 \bc \label{mp}(Cf. \cite{Ma69}) \\ 

(i) $\nabla\,\Ann(s)^\times\subseteq\{\varphi' \lambda\,_1-\varphi \lambda\,_1'\,:\  \varphi'\neq 0, \ |\varphi |\leq |\varphi'|-e_s\}\subseteq\Ann(s)^\times$;\\

(ii)   $\nabla\,\Min(s)\subseteq\{\varphi' \lambda\,_1-\varphi \lambda\,_1'\,:\ \varphi' \in \D^\times,\ |\varphi |\leq -e_s\}\subseteq\Min(s)$;\\

(iii) if $2\LC\leq n$ then $\nabla\,\Min(s)\subseteq\{\varphi' \lambda\,_1\,:\ \varphi' \in \D^\times\}\subseteq\Min(s).$\\

Moreover if $\nabla$ is a unit of $\D$ (for example if $\D$ is a field) the inclusions are equalities.
\ec
\bpr (i) If $f_1\in \Ann(s)^\times$ then $\nabla f_1=m'\lambda\,_1-m\lambda\,'_1$ where $|m'|=|f_1|-\LC$ and $|m|\leq |f_1|+\LC-n-1=|m'|-e_s$ by Theorem \ref{annconverse}. If $\varphi'\neq 0$ and $|\varphi |\leq |\varphi'|-e_s$ then $\varphi' \lambda\,_1-\varphi \lambda\,_1'\in\Ann(s)^\times$ by Lemma \ref{generalann}. (ii) If $f_1\in\Ann(s)$ and $|f_1|=\LC$ then by Theorem \ref{annconverse}, $|m'|=0$. If $f_1=\varphi'\lambda_1-\varphi\lambda_1'$, $\varphi'\in\D^\times$  and $|\varphi|\leq -e_s$ then $f_1\in\Ann(s)$ and $|f_1|=|\lambda_1|=\LC$ by Lemma \ref{generalann}. Part (iii) is an immediate consequence of (ii). If $f_1\in\Ann(s)^\times$ and $\nabla$ is a unit of $\D$ then $f_1/\nabla\in\Ann(s)^\times$ and hence $f_1\in\nabla\Ann(s)^\times$.
\epr
Thus if  $e_s\leq 0$ we have $\lambda_1-\lambda'_1\in\Min(s)$, as is well-known for sequences over a field. From Theorem \ref{simplicityitself} and Corollary \ref{mp} we have 
\bc \label{mp2}(Cf. \cite[Theorem 1]{Nied87}) Let $S$ be an infinite sequence over $\F$, $n\in[n_i,n_{i+1})$ and $q=q^{(i)}$, $q'=q^{(i-1)}$. If $s=S|n$ is essential  then 
$$\Ann(s)^\times=\{\varphi' q_1-\varphi q_1':  \varphi'\neq 0, \ |\varphi |\leq |\varphi'|-e_s\}, \Min(s)=\{\varphi' q_1-\varphi q_1':\varphi' \in \F^\times,\ |\varphi |\leq -e_s\}.$$
\ec

\subsubsection{Solutions}\label{esoln}
Next we look at solutions i.e. pairs $(f_1,f_2)$ with $f_1\in\Ann(s)^\times$ and $x\,f_2=[f_1\cdot\ul{s}]$.
 \bl\label{phiphi'}Let $\varphi,\varphi'\in\R$. If $g_1=\varphi'\lambda\,_1-\varphi\lambda\,'_1$, $0\leq |\varphi'|\leq n-\LC$ and  $|\varphi|\leq \LC-1$ then (i) $g_2=
 \varphi' \lambda\,_2-\varphi \lambda\,'_2$\,; (ii) $\nabla\, \varphi=\langle g,\lambda\,\rangle$ and  $\nabla\, \varphi'=\langle g,\lambda\,'\rangle$.
 \el
 \bpr (i) We have $\lambda\,_1\cdot\ul{s}=M+x\,\lambda\,_2$ and $\lambda\,'_1\cdot\ul{s}'=M'+x\,\lambda\,_2'$ where $\vv(M)\leq \LC-n$ and $\vv(M')\leq \LC'-n'=1-\LC$. 
 Write $\ul{s}=(\ul{s}-\ul{s}')+\ul{s}'$ so that $\vv(\ul{s}-\ul{s}')\leq -n'$ and put $N'=\lambda\,_1'\cdot(\ul{s}-\ul{s}')$.  This gives
 \begin{eqnarray*}g_1\cdot\ul{s}&=&(\varphi'\lambda\,_1-\varphi\lambda\,'_1)\cdot\ul{s}
 =\varphi'\cdot(M+x\,\lambda\,_2)-\varphi\lambda\,_1'\cdot((\ul{s}-\ul{s}')+\ul{s}')\\
&=&\varphi'\cdot(M+x\,\lambda\,_2)-\varphi\cdot N'-\varphi\cdot(M'+x\,\lambda\,_2')\\
&=&\varphi' \cdot M -\varphi\cdot N'-\varphi\cdot M'+x\,(\varphi'\lambda\,_2-\varphi\lambda\,'_2).
\end{eqnarray*}
Further, $\vv(\varphi'\cdot M)\leq |\varphi'|+\LC-n\leq 0$ by  hypothesis and similarly  $\vv(\varphi\cdot M')\leq 0$. Now
$$\vv(\varphi\cdot N')=|\varphi|+|\lambda\,_1'|+\vv(\ul{s}-\ul{s}')
  \leq |\varphi|+\LC'-n'=|\varphi|+1-\LC\leq 0$$ as
 $ |\varphi|\leq \LC-1$.
  Thus  $\vv(\varphi'\cdot M -\varphi\cdot N'-\varphi\cdot M')\leq 0$ and 
$g_2=\varphi'\lambda\,_2-\varphi\lambda\,'_2$. (ii) We have
\begin{eqnarray*}\langle g,\lambda\,\rangle&=&
g_2\,\lambda\,_1-g_1\,\lambda\,_2=(\varphi'\,\lambda\,_2-\varphi\,\lambda\,'_2)\,\lambda\,_1-(\varphi'\,\lambda\,_1-\varphi\,\lambda\,'_1)\lambda\,_2\\
&=&\varphi\,(\lambda\,_2\,\lambda\,_1'-\lambda\,_1\,\lambda\,'_2)=\nabla \varphi.
\end{eqnarray*}
Similarly
$\langle g,\lambda\,'\rangle=
g_2\,\lambda\,'_1-g_1\,\lambda\,'_2=\varphi'\,(\lambda\,_2\,\lambda\,_1'-\lambda\,_1\,\lambda\,'_2)=\nabla \varphi'$.
\epr

\bc\label{andtheresmore} Let $f$ be a solution for $s$, $|f_1|\leq n$ and $m'=\langle f,\lambda\,'\rangle$, $m=\langle f,\lambda\,\rangle$. 

(i) (uniqueness) If $\nabla\, f_1=\varphi'\lambda\,_1-\varphi\lambda\,'_1$ where $0\leq |\varphi'|=|f_1|-\LC$ and $|\varphi|\leq \LC-1$ then  $\varphi=m$ and $\varphi'=m'$\,; 
 
(ii)  $\nabla f_2=m'\lambda_2\,-m \lambda_2'$\,; 

(iii) (degree bound) if $\lambda\,'_2\neq 0$ then $|m|+|\lambda\,_2'|\leq |f_2|-1$ and $|f_2|=|m'|+|\lambda\,_2|$.
\ec
\bpr (i) Applying Lemma \ref{phiphi'} to $g_1=\nabla\, f_1=\varphi'\lambda\,_1-\varphi\lambda\,'_1$ gives $\nabla\, m=\nabla \langle f,\lambda\,\rangle=\langle \nabla \,f,\lambda\,\rangle=\langle g,\lambda\,\rangle=\nabla\varphi$. Therefore $\varphi=m$ and $\varphi'=m'$.

(ii)   We know from Theorem \ref{annconverse} that $\nabla\, f_1=m'\,\lambda\,_1-m\,\lambda\,'_1$ where $|m'|\leq n-\LC$ and $|m|\leq \LC-1$. Hence Lemma \ref{phiphi'} implies that $\nabla f_2=m' \lambda\,_2-m\lambda\,'_2$.

 (iii)  We have $\lambda\,'\in\Ann(s')$ and if $s'$ is trivial then $\lambda\,'=(c,0)$ for some $c\in\D^\times$. As $\lambda\,'_2\neq 0$,  $s'$ is non-trivial so $1-n'\leq \vv(s')\leq 0$ and therefore $\vv(s')=\vv$. From Theorem \ref{annconverse}, $|m|+\LC'\leq |f_1|-1$, so $$|m|+|\lambda\,'_2|=|m|+(\vv(s')+|\lambda\,'_1|-1)= |m|+\vv+\LC'-1\leq |f_1|-1+\vv-1=|f_2|-1.$$
From Part (ii) we have $\nabla f_2=m'\lambda\,_2-m \lambda\,'_2$, so  $|f_2|=|m'|+|\lambda\,_2|$.
\epr
The final result of this section on solutions is a simple consequence of Corollary \ref{andtheresmore}.
 \bc \label{mpsol}(Cf. \cite{Ma69}) If $n\geq 2$, $s=s_0,\ldots,s_{1-n}$ is an essential sequence and $\Sigma$ denotes the solutions $\{(f_1,f_2): f_1\in\Ann(s),\, 0\leq |f_1|\leq n\}$ then  
$$
\nabla\,\Sigma\subseteq\{\varphi'\, \lambda-\varphi\, \lambda',\, 0\leq|\varphi'|\leq n-\LC, \ |\varphi |\leq |\varphi'|-e_s\}\subseteq\Sigma
$$
and if $\nabla$ is a unit of $\D$ (for example if $\D$ is a field) the inclusions are equalities.
\ec
We leave the corresponding result for minimal solutions to the interested reader.
\section{Some Applications of Decomposition}
We give some applications of the results from the previous sections. As usual, $n\geq1$, $s=s_0,\ldots,s_{1-n}$ is non-trivial and $\mu,\mu'$ are  obtained using Algorithm \ref{calPA} or, if $\D$ is a field $\F$, using the Normalised Algorithm \ref{calPA}. We put $\LC=\LC(s)$.
\subsection{Sequences over a Field}
We prove several    gcd-related results,  relate partial quotients to $\mu_1,\mu'_1$  and count  the number of solutions  when $|\F|<\infty$. Firstly a partial converse to Proposition \ref{minfield}(ii). 
\bc \label{minfieldconverse}  If $2\LC\leq n$, $f_1\in\Ann(s)^\times$ and $ |f_1|\leq n-\LC$  then

(i)   $\nabla_s\,f=m'\mu$;

(ii) if in addition $\gcd(f_1,f_2)=1$ then $m'\in\F^\times$ i.e. $f_1\in\Min(s)$.
\ec
\bpr  (i)   Since $\LC\leq n-\LC$, such an $f_1$ can exist. Proposition \ref{PSGannconverse} or  Theorem \ref{annconverse} imply that $\nabla_s\,f_1=m'\,\mu_1-m\,\mu'_1$ where $|m'|=|f_1|-\LC\geq 0$ and $|m|\leq |f_1|+\LC-n-1\leq -1$, so $\nabla_s\, f_1=m'\mu$. Since $|f_1|\leq n$, we also have $\nabla_s\,f_2=m'\,\mu_2-m\,\mu'_2$ from Corollary \ref{1mp} or Corollary \ref{andtheresmore} i.e. $\nabla_s\,f=m'\mu$.  (ii)  By Corollary \ref{gcdcor}, $\gcd(\mu_1,\mu_2)=1$, so $\nabla_s=\gcd(\nabla_s\, f_1,\nabla_s\, f_2)=\gcd(m'\mu_1,m'\mu_2)=m'$. Thus $m'\in\F^\times$ and $|f_1|=|\mu_1|=\LC$.
\epr
The example after Proposition \ref{minfield}  shows that the condition $|f_1|\leq n-\LC$ is necessary. Secondly, one may show directly that if $f_1,g_1\in\Ann(s)^\times$ and $|f_1|+|g_1|\leq n$ then $\langle f,g\rangle =0$; see \cite[Corollary 3.25]{N95b}. This gives another proof of Corollary \ref{minfieldconverse}.
\bc \label{lrsappl}(Cf. \cite[p.  439-444]{LN83}). Let  $S$ be a linear recurring sequence over $\F$, $\Id_S=g_1\,\F[x]$ where $g_1$ is monic, $n\geq 2|g_1|$ and $s=S|n$. Then 

(i) $g$ is a minimal solution for $s$ and any minimal solution of $s$ is $c\,g$ for some $c\in\F^\times$; 

(ii) $\ul{S}=[g_1\cdot \ul{s}]/g_1$ and $S$ is determined by $S_0,\ldots,S_{2|g_1|-1}$;

(iii) if $i$ is the first index such that $b_{i+1}=0$ in obtaining the partial quotients of  $\ul{S}$, then $g=q^{(i)}/\lc(q_1^{(i)})$.

\ec
\bpr  (i) Firstly, $g$ is a solution for $s=S|n$. As $g_1$ is a minimal polynomial of $S$, $\gcd(g_1,g_2)=1$  and $xg_2=[g_1\cdot \ul{S}]=[g_1\cdot \ul{s}]$ by Lemma \ref{Ann(S|n)} since $n\geq |g_1|$. As $|g_1|\leq n-|g_1|$, $g_1$  is a minimal solution for $s$ by Corollary \ref{minfieldconverse}.  Further $e_s>0$ so any minimal solution for $s$ is $c\,g$ where $c\in\F^\times$ by Corollary \ref{1mp} or \ref{mp}. (ii) We have $\ul{S}=[g_1\cdot \ul{S}]/g_1=[g_1\cdot \ul{s}]/g_1=xg_2/g_1$.  Since $g_1$ is  uniquely determined by $s_0,\ldots,s_{2|g_1|-1}$,  so are  $g_2$ and $S$. (iii) If $s$ is geometric, this is Example \ref{geometric}. Suppose that  $s$ is essential and put $q=q^{(i)}$, $q'=q^{(i-1)}$. We have $n_{i+1}=\infty$ since $\ul{S}\in\F(x)$ and $n_{i}=|q_1'|+|q_1|<2|q_1|=2\LC\leq n<n_{i+1}$. From Theorem \ref{simplicityitself}, $q$ is a minimal solution for $s$ and $g$ is the unique monic solution of $s$ by Corollary \ref{mp} since $e_s>0$ and $\LC\leq2\LC\leq n$.
\epr
Instances of Part(iii) of Corollary \ref{lrsappl} were given in Example \ref{WSexample2}.

\bc\label{divisors}   Suppose that $f$ is a  solution for $s$ such that $|f_1|\leq n$ and let $m=\langle f,\mu\rangle$, $m'=\langle f,\mu'\rangle$. Then $\gcd(m,m')=\gcd(f_1,f_2)$.
\ec
\bpr By definition, $m=f_2\mu_1-f_1\mu_2$ and $m'=f_2\mu_1'-f_1\mu_2'$ so that if  $d\,|f_1,f_2$ then $d\,|m,m'$.  We also know that $\nabla f_1=m'\mu_1-m\mu_1'$ by  Proposition \ref{identity}.    Corollary \ref{andtheresmore} implies that $\nabla f_2=m'\mu_2-m\mu_2'$ since $|f_1|\leq n$. Hence if $d\,|m,m'$ then $d\,|f_1,f_2$. 
\epr

For the next result, $\F_q$ is a finite field with $q<\infty$ elements. If $d\geq 0$,  the number of polynomials with coefficients in $\F_q$ of degree $d$ is $N_d=(q-1)q^d$ and the number of polynomials of degree at most $d$ is $1+\sum_{k=0}^{d}N_k$. Results of Section \ref{decomp1} now easily give the number of solutions for $s$ with denominator of degree $d$ when $\LC\leq d\leq n$:
\bc\label{Anncount}  For $\LC\leq d\leq n$, the number of solutions for $s$ with denominator of degree $d$ is $N_{d-\LC}\left(1+\sum_{k=0}^{d-\LC-e_s}\right).$
\ec
\bpr Let $E_d=\{(f_1,f_2): f_1\in\Ann(s)^\times: |f_1|=d\}$. From Corollary \ref{1sol} or Corollary \ref{mp},  we have $f\in E_d$ if and only if $f_1=\varphi'\mu_1-\varphi\mu'_1$ where (i) $|\varphi'|=d-\LC$ and (ii) $\varphi=0$ or $|\varphi|\leq d-\LC-e_s$\,, which yields the stated result.
\epr

\subsection{Non-Vanishing Annihilating Polynomials}
We consider the following problem: let $a\in\D$ be arbitrary and suppose that $\mu_1(a)=0$. Find a solution $\xi=(\xi_1,\xi_2)$ such that $\xi_1(a)\neq 0$  and $\xi_1$ has least degree among solutions with first component not vanishing at $a$. 
We begin with a pseudo-geometric  example.

\be\label{PSGSd}
Let $n\geq 2$ and $s=s_0,\ldots,s_{1-n}=1,0^{n-1}$. Then $e_s=n-1>0$ and $\nabla_s=1$, so $\Min(s)=\{\varphi'\,x: \varphi'\in\D^\times\}$  by Corollary \ref{1mp}. Thus all minimal polynomials of $s$ vanish at 0. However $g=(x^n+1,x^{n-1})$ is a solution for $s$ and $g_1(0)\neq0$. We will shortly see that $\min\{|f_1|:\ f\mbox{ is a solution for }s, f_1(0)\neq 0\}=n$, so that $|g_1|$ attains this minimum. 
\eex

 We can assume that $s$ is non-trivial and $\LC\geq 1$, for otherwise $\mu_1\in\D^\times$ vanishes nowhere. 
Put  $\Ann(s)^{(a)}=\{f_1\in\Ann(s):\  f_1(a)\neq 0\}$.   Any polynomial of degree $n$ which does not vanish at $a$ annihilates $s$, so that $$\LC^{(a)}=\min\{|f_1|:\ f_1\in\Ann(s)^{(a)}\}$$
is well-defined and $\LC\leq \LC^{(a)}\leq n$. We put $\Min(s)^{(a)}=\{f_1\in\Ann(s)^{(a)}:  |f_1|=\LC^{(a)}\}$. Example \ref{PSGSd} shows that $\LC^{(a)}-\LC$ can be arbitrarily large. As usual, $\mu'$ is obtained as in Algorithm \ref{calPA}.
\bc \label{nonzero}   If  $\mu_1(a)=0$ then   $\mu_1'(a)\neq 0$.
\ec
\bpr 
Proposition \ref{identity} yields $\mu_2\mu_1'-\mu_1\mu'_2\in\F^\times$, so $\mu_1'(a)\neq 0$ (and $\mu_2(a)\neq 0$). 
\epr

Using  Proposition \ref{PSGannconverse}, Theorem \ref{annconverse} and Corollary \ref{nonzero}, we can now solve the problem posed at the head of this subsection.
\bt \label{myversion}(Cf. \cite[Proof of Theorem 3.7]{Salagean} ) Let $n\geq 1$, $s_0,\ldots,s_{1-n}$ be a sequence over $\D$,  $e=n+1-2\LC_n$ and $M=\max\{e,0\}$. If $\mu_1(a)=0$ then $\LC^{(a)}=\LC+M$. In fact  $\xi_1=x^M\mu_1-\mu'_1\in\Min(s)^{(a)}$  and  $\xi_2=x^M\mu_2-\mu_2'$.
\et
\bpr  If $n=1$ and $\mu_1(a)=0$ then $a=0$ and $e=0$; $\xi_1=\mu_1-\mu_1'=x-1\in\Min(s)$ satisfies $\xi_1(a)\neq 0$ and $\xi_2=s_0=\mu_2-\mu_2'$, so $\xi$ is the required solution and $\LC^{(a)}=\LC+M$. 

If $n\geq 2$ and 
 $e\leq 0$ then $\xi_1=\mu_1-\mu_1'\in\Min(s)$  so $\LC=\LC^{(a)}$ and $\xi_2=\mu_2-\mu_2'$  by Lemma \ref{phiphi'}.  Also $\xi_1(a)\neq 0$   by Corollary \ref{nonzero} and so $\xi$ is the required solution. 
 
 Now let $n\geq 2$ and $e>0$. We have 
  $\LC+e=n+1-\LC\leq n$ since $\LC\geq 1$. We first show that $\LC^{(a)}\geq \LC+e$. Let $f_1\in\Ann(s)^{(a)}$, $f=(f_1,f_2)$, $m=\langle f,\mu\rangle$ and $m'=\langle f,\mu'\rangle$. From Proposition \ref{PSGannconverse} or Theorem~\ref{annconverse} we have $\nabla_s\, f_1=m'\mu_1-m\mu'_1$ where $|m'|=|f_1|-\LC\geq 0$ and $|m|\leq |f_1|+\LC-n-1$.  If  $|f_1|+\LC-n-1<0$ then $m=0$, $\nabla_s\, f_1(a)=m'(a)\,\mu_1(a)=0$ and $f_1\not\in\Ann(s)^{(a)}$ for a contradiction. Hence $|f_1|\geq n+1-\LC=\LC+e$ and  $\LC^{(a)}\geq \LC+e=\LC+M$. 
  
To see that $\LC^{(a)}\leq \LC+M$, let $\xi_1=x^M\mu_1-\mu_1'$ which has degree $\LC+M$. We have $\xi_1(a)=-\mu_1'(a)\neq 0$ by Corollary \ref{nonzero}. We claim that $\xi_1\in\Min(s)^{(a)}$. We have $|1|=0=|\xi_1|+\LC-n-1$, so by Lemma \ref{PSGSann} or Lemma \ref{generalann} we have $\xi_1\in\Ann(s)^{(a)}$ and $\LC^{(a)}\leq|\xi_1|=\LC+M$. 

Finally, we verify that $\xi_2=x^M\mu_2-\mu'_2$. Put $\varphi'=x^M$ and $\varphi=1$. If $s$ is pseudo-geometric,  Lemma \ref{PSGphiphi'} applies since $|\varphi|=0\leq \LC-1$  and $0\leq M\leq n-1$.  Hence $\xi_2=x^M\mu_2-\mu'_2=x^M\mu_2$. Suppose that $s$ is essential. We have $0\leq M\leq n+1-2\LC\leq n-\LC$ since $\LC\geq 1$, so that Lemma   \ref{phiphi'} applies and $\xi_2=x^M\mu_2-\mu'_2$ in this case too.
\epr
 Theorem \ref{myversion} yields the following simple extension of Algorithm \ref{calPA}. 

\begin{algorithm}\label{calPA+} (Cf. \cite[Algorithm 3.2]{Salagean})

\begin{tabbing} {\tt Input}:  $n\geq 1$, $a\in \D$ and   sequence $s=s_0,\ldots,s_{1-n}$ over $\D$.\\
{\tt Output}:\   Solution $\xi$ for $s$ such that $\xi_1\in\Min(s)^{(a)}$.\\\\
$\lceil$ {\tt Algorithm} \rm{\ref{calPA}}\ $({\tt input}: n,s; {\tt output}: \mu,\mu')$; \\\\

{\tt if}\=\ $\mu_1(a)\neq 0$ \={\tt then} $\xi\leftarrow \mu$ {\tt else}\ $\xi\leftarrow x^{\max\{n+1-2|\mu_1|,\,0\}}\mu-\mu';$\\\\
 {\tt return} $\xi$.$\rfloor$
\end{tabbing}
\end{algorithm}
\begin{table}\label{shortex}
\caption{Algorithm \ref{calPA} for $0,1,1,0, 0,1,0,1$ over $\F_2$}
\begin{center}
\begin{tabular}{|l| c| c| l| l|}\hline
$s$ & $\Delta_1$  & $e_s$ &$\mu$ &$\mu'$\\\hline\hline
$\ $   &$-$  &$- $  & $(1,0)$ & $(0,1)$  \\\hline
$0$   &$0$  &$1$  & $(1,0)$ &$(0,1)$ \\\hline
$0,1$   & $1$ &$2$ & $(x^2,1)$ & $(1,0)$  \\\hline
$0,1,1$   & $1$ &$-1$  & $(x^2+x,1)$ & $(1,0)$  \\\hline
$0,1,1,0$   &$1$  &$0$  & $(x^2+x+1,1)$ & $(1,0)$  \\\hline
$0,1,1,0, 0$   & $1$ &$1$  & $(x^3+x^2+x+1,x)$  & $(x^2+x+1,1)$ \\\hline
$0,1,1,0, 0,1$   &$0$  &$0$  & $(x^3+x^2+x+1,x)$ &  $(x^2+x+1,1)$ \\\hline
$0,1,1,0, 0,1,0$   & $1$ & $1$ & $(x^4+x^3+1,x^2+1)$ & $(x^3+x^2+x+1,x)$ \\\hline
$0,1,1,0, 0,1,0,1$   &$1$  & $0$ & $(x^4+x^2+x,x^2+x+1)$ &  $(x^3+x^2+x+1,x)$.\\\hline
\end{tabular}
\end{center}
\end{table}

 \begin{example} \label{bullet} Table 2 gives the iterations of Algorithm \ref{calPA} (implemented  using \cite{COCOA}) for the  sequence $s$ over $\F_2$ of \cite[Table I]{Salagean}; we have omitted $\Delta'_1$ as it is the constant $1$.  We see that $s$ is essential, $e=e_s=1$ and $\mu_1(0)=0$. Theorem \ref{myversion} implies that $\LC^{(0)}=\LC+e=5$ and  Algorithm \ref{calPA+} gives $\xi=x^e\,\mu+\mu'=(x^5+x+1,x^3+x^2)$. 
\end{example}

\brs 
(i) Algorithm \ref{calPA+} is  simpler than \cite[Algorithm 3.2]{Salagean},  e.g. it does not include  tests on $\mu_1'$\,. It also computes $\mu_2$.  Corollary \ref{factorialchar}(ii) below, a version of  which was stated without proof in \cite{Salagean}, was used to justify Algorithm 3.2, {\em loc. cit.} In Algorithm 3.2, {\em loc. cit.} the polynomial $\mu_1'$ is initialised to $1$ as in \cite{Ma69} rather than $0$; see Remark \ref{arbyd}.

(ii)  The original motivation of  \cite{Salagean}: let $a=0$ and $\xi_1^\ast$ be the reciprocal of $\xi_1$.  Since $\xi_1(0)\neq 0$, $|\xi_1^\ast|=|\xi_1|=\delta$ say,  
$(\xi_1^*\cdot\ul{s}^\ast)_i= (\xi_1\cdot\ul{s})_j$
where $j=1-n+\delta-i$, and $\delta+1-n\leq i\leq 0$ if and only if $\delta+1-n\leq j\leq 0$. Hence if $1\leq \delta<n$, $\xi_1$ and the first $\delta$ terms $s_0\ldots,s_{1-\delta}$ uniquely determine the last $n-\delta$ terms $s_{-\delta},\ldots,s_{1-n}$ if and only if $1\leq n-\delta<n$, $\xi_1^\ast$ and the last $n-\delta$ terms $s_{1-n}\ldots,s_{\delta-n}$ uniquely determine the first $\delta$ terms $s_{\delta-n+1},\ldots,s_0$.
\ers
We can also construct an element of $\Min(s)^{(a)}$ by extending  $s$ by one term.
\bc \label{dge0}(Cf. \cite{Salagean})  Let  $s=s_0,\ldots,s_{1-n}$, $\mu_1\in\Min(s)$ and $\mu_1(a)=0$. Suppose that $e=e_s\geq 1$. Put $t=s,s_{-n}$ where $s_{-n}$ is chosen so that $\Delta(\mu_1;t)\neq 0$.
If $\nu_1$ is obtained as in Theorem \ref{indthm} then $\nu_1\in\Min(s)^{(a)}$.
\ec
\begin{proof}  Let $\Delta_1=\Delta(\mu_1;t)$. Since $e\geq 1$,  $\nu_1=\Delta'_1\, x^e\mu_1-\Delta_1\,\mu_1'\in \Min(t)$ from Theorem \ref{indthm} and $|\nu_1|=n+1-\LC=\LC^{(a)}$ by Theorem \ref{myversion}. Further,
$\nu_1(a)=-\Delta_1\,\mu_1'(a)\neq 0$ by Corollary \ref{nonzero} and since $\Ann(t)\subseteq\Ann(s)$, $\nu_1\in \Min(s)^{(a)}$. 
\epr

For the Example of Table 2,  $\Delta_9=1$ requires $s_9=0$ and we obtain $\nu=\xi$ as before. 
\bc \label{factorialchar}Let $s=s_0,\ldots,s_{1-n}$ be a sequence over $\F$, $\mu,\mu'$ be as usual and  $m=\langle f,\mu\rangle\,,m'=\langle f,\mu'\rangle$.  If  $a\in \F$, $\mu_1(a)=0$ and $M=\max\{e_s,0\}$ then
\begin{tabbing}
(i)\ \ \= $\Ann(s)^{(a)}=\{\varphi'\mu_1-\varphi\mu'_1, |\varphi'|\neq 0, |\varphi|\leq |\varphi'|-e_s, \varphi(a)\neq 0\}$;\\\\
(ii) \>$\Min(s)^{(a)}$\ \= =\ $\{f_1\in\Ann(s)^{(a)}: |m'|= M, |m|\leq|f_1|+\LC-n-1,m(a)\neq 0\}$;\\\\
\>\>=\ $\{\varphi'\mu_1-\varphi\mu_1':\ |\varphi'|= M,  |\varphi|\leq M-e_s, \varphi(a)\neq 0 \}$.
\end{tabbing}
\ec
\bpr  (i) This is a restatement of Corollary \ref{1mp} or  Corollary \ref{mp}. (ii) We have $\nabla f_1=m'\mu_1-m\mu_1'$ and  $f_1\in\Ann(s)$ if and only if $ |m'|=|f_1|-\LC$ and $ |m|\leq|f_1|+\LC-n-1$. Also $|f_1|=\LC+M$ from Theorem \ref{myversion} since $f_1\in\Min(s)^{(a)}$. Similarly if $g_1=\varphi'\mu_1-\varphi\mu_1'$ then $g_1\in\Ann(s)$ if and only if  $|\varphi'|=|g_1|-\LC$, $|\varphi|\leq |\varphi'|-e_s$ by Lemma \ref{PSGSann} or Lemma \ref{generalann} and  $|g_1|=\LC+M$ by Theorem \ref{myversion}.
\epr

For Example \ref{bullet}, $e_s=1$ and   so 
$\Min(s_0,\ldots,s_{-7})^{(0)}=\{x\,\mu_1+\mu_1', (x+1)\mu_1+\mu_1'\}$ .


\begin{thebibliography}{10}

\bibitem{Be68}
{Berlekamp, E. R.}
\newblock {\em Algebraic Coding Theory}.
\newblock Series in Systems Science. McGraw Hill, New York-Toronto, 1968.

\bibitem{Cheng}
{Cheng, U.}
\newblock {On the Continued Fraction and Berlekamp's Algorithm}.
\newblock {\em IEEE Transactions on Information Theory}, 30:541--544, 1984.

\bibitem{COCOA}
{Cocoa Team}.
\newblock {\em {A System for Doing Computations in Commutative Algebra}}.
\newblock {Available at http://cocoa.dima.unige.it}, Version 5.0.

\bibitem{Gus76}
{Gustavson, F. G.}
\newblock Analysis of the {Berlekamp-Massey} linear feedback shift-register
  synthesis algorithm.
\newblock {\em IBM J. Res. Dev.}, 20:204--212, 1976.

\bibitem{LN83}
{Lidl, R. and Niederreiter, H.}
\newblock{\em Finite Fields. Encyclopedia of 
                  Mathematics and its Applications}.
\newblock{Addison-Wesley, Reading}, 20, 1983.

\bibitem{McWS}
{MacWilliams, F. J. and Sloane, N. J. A.}
\newblock {\em {The Theory of Error-Correcting Codes}}.
\newblock North Holland, Amsterdam, 1977.

\bibitem{Ma69}
{Massey, J. L.}
\newblock {Shift-Register Synthesis and {BCH} Decoding.}
\newblock {\em IEEE Trans. Inform. Theory}, 15:122--127, 1969.

\bibitem{Mills}
{Mills, W. H.}
\newblock {Continued Fractions and Linear Recurrences.}
\newblock {\em {Mathematics of Computation}}, 29:173--180, 1975.

\bibitem{Nied87}
{Niederreiter, H.}
\newblock {Sequences with Almost Perfect Linear Complexity Profile. Advances in
  Cryptology --- Eurocrypt '87 (D. Chaum, W.L. Price, Eds.)}.
\newblock {\em {Lecture Notes in Computer Science}}, 304:37--51, 1987.

\bibitem{Northcott}
{Northcott, D. G.}
\newblock {Injective Envelopes and Inverse Polynomials}.
\newblock {\em {J. London Math. Soc.}}, 8:290--296, 1974.

\bibitem{N95b}
{Norton, G. H.}
\newblock {On the Minimal Realizations of a Finite Sequence}.
\newblock {\em J. Symbolic Computation}, 20:93--115, 1995.

\bibitem{N99b}
{Norton, G. H.}
\newblock {On Shortest Linear Recurrences}.
\newblock {\em J. Symbolic Computation}, 27:323--347, 1999.

\bibitem{N10a}
{Norton, G. H.}
\newblock {The Berlekamp-Massey Algorithm via Minimal Polynomials.}
\newblock {\em {math.ArXiv: 1001.1597}}, pages 1--22, 2010.

\bibitem{Salagean}
{Salagean, A.}
\newblock {An Algorithm for Computing Minimal Bidirectional Linear Recurrence
  Relations.}
\newblock {\em {IEEE Trans. Info. Theory}}, 55:4695--4700, 2009.

\bibitem{WS}
{Welch, L. R. and Scholtz, R. A.}
\newblock {Continued Fractions and Berlekamp's Algorithm.}
\newblock {\em {IEEE Trans. on Information Theory}}, 46:19--27, 1979.

\end{thebibliography}
\end{document}